\documentclass{cernrep} 

\usepackage{fancyhdr}
\usepackage[colorlinks=true, linkcolor=black, citecolor=black, filecolor=black, urlcolor=blue]{hyperref}
\fancyhfoffset{4 mm}
\fancypagestyle{ARTTITLE}{%
\fancyhf{} 
\lhead{\small{Proceedings of the CERN--Accelerator--School course: \it{Introduction to Accelerator Physics}}}
\lfoot{Available online at \url{https://cas.web.cern.ch/previous-schools}}
\rfoot{\thepage\hspace*{3mm}}

}

\usepackage{texnames}
\usepackage[T1]{fontenc}

\usepackage{texnames}
\usepackage[T1]{fontenc}

\usepackage{afterpage}
\def\unsty{\ \mathrm}

\newcommand{\tabitem}{~~\llap{\textbullet}~~}
\newcommand{\phantomtabitem}{~~\phantom{\llap{\textbullet}}~~}

\begin{document}
\title{RF Systems} 
 
\author {H. Damerau}

\institute{CERN, Geneva, Switzerland}

\begin{abstract}
    Radio-frequency (RF) systems deliver the power to change the energy of a charged particle beam, and they are integral parts of linear and circular accelerators. A longitudinal electrical field in the direction of the beam is generated in a resonant structure, the RF cavity. As it directly interacts with the bunches of charged particles, the cavity can be considered as a coupler to transport energy from an RF power power amplifier to the beam. The power amplifier itself is driven by a low-level RF system assuring that frequency and phase are suitable for acceleration, and feedback loops improve the longitudinal beam stability. The spectrum of RF systems in particle accelerators in terms of frequency range and RF voltage is wide. Special emphasis is given to the constraints and requirements defined by the beam, which guides the appropriate choices for the RF systems.
\end{abstract}


\maketitle 
\thispagestyle{ARTTITLE} 

\section{Introduction}

\begin{figure}[htb]
	\centering
	\includegraphics[width=0.8\linewidth]{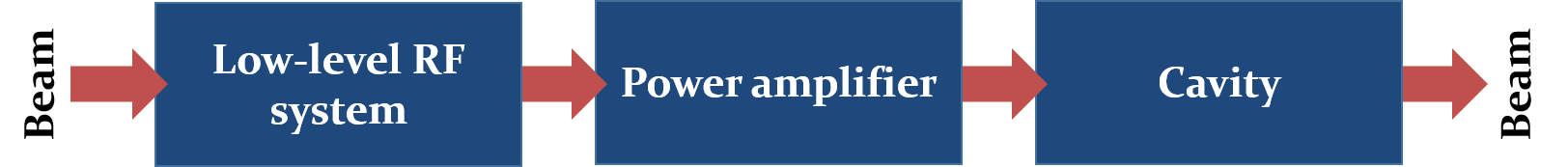}
	\caption{RF system overview.}
	\label{fig:RFSystemOverviewBeamToBeam}
\end{figure}

RF systems provide the longitudinal electric fields for the energy increase or decrease of charged particles and transform a string of magnets into an accelerator. These high electric fields are conveniently achieved by resonantly exciting an oscillating system, and the accelerating voltage becomes
\begin{equation*}
    V (t) = V_0 \sin ( 2 \pi f t + \phi) \, ,
\end{equation*}
where $V_0$ is the peak voltage amplitude, $f$ the frequency and $\phi$ the RF phase. At the same time the voltage generated by the RF system keeps the particles longitudinally focused in bunches, and it allows to manipulate the longitudinal beam parameters. The main ingredients of a typical RF system are sketched in Fig.~\ref{fig:RFSystemOverviewBeamToBeam}.
The cavity, the oscillator in which the large accelerating fields are built up is the most visible part of an RF system and it directly interacts with the beam. Of equal importance are the power amplifiers. They produce the power which is converted into an energy change of the beam. The amplifiers are driven by the low-level RF system providing the necessary small-amplitude signals at the correct frequency and phase. In many accelerators these are derived from the beam itself. 

The first part of this overview concentrates on the choice of the basic RF system parameters, driven by the constraints and requirements defined by the beam. Thereafter, the RF cavity as a key component is treated from conceptual and engineering points of view. The main technologies for power amplifiers are described, together with guidelines to select the most appropriate type for a given application. The basic feedback systems and beam control loops are introduced to generate RF signals at the correct amplitude and phase for the acceleration in a synchrotron.

\section{Choice of parameters}

RF systems in particle accelerator exist in a wide frequency and voltage range. While the frequency can be as low as few hundred kilohertz for proton synchrotrons in the low-energy range~(up to a few $\mathrm{GeV}$ for protons), the original RF frequency for the Compact Linear Collider (CLIC) used to be as high as $30\unsty{GHz}$, which had later been changed to its present value of $12\unsty{GHz}$. Also the amplitude of signals in RF systems spans many orders of magnitude. RF signals from beam pick-ups may be as low few microvolts. At the upper end of the range, the total RF voltage available in the Large Electron Positron (LEP) collider at CERN was about $3.6\unsty{GV}$ just before the final shut-down of the accelerator in the year 2000. Planned high-energy linear accelerators will require pulsed RF voltages in the teravolt range. 

\begin{figure}[htb]
    \centering
	\includegraphics[width=0.7\linewidth]{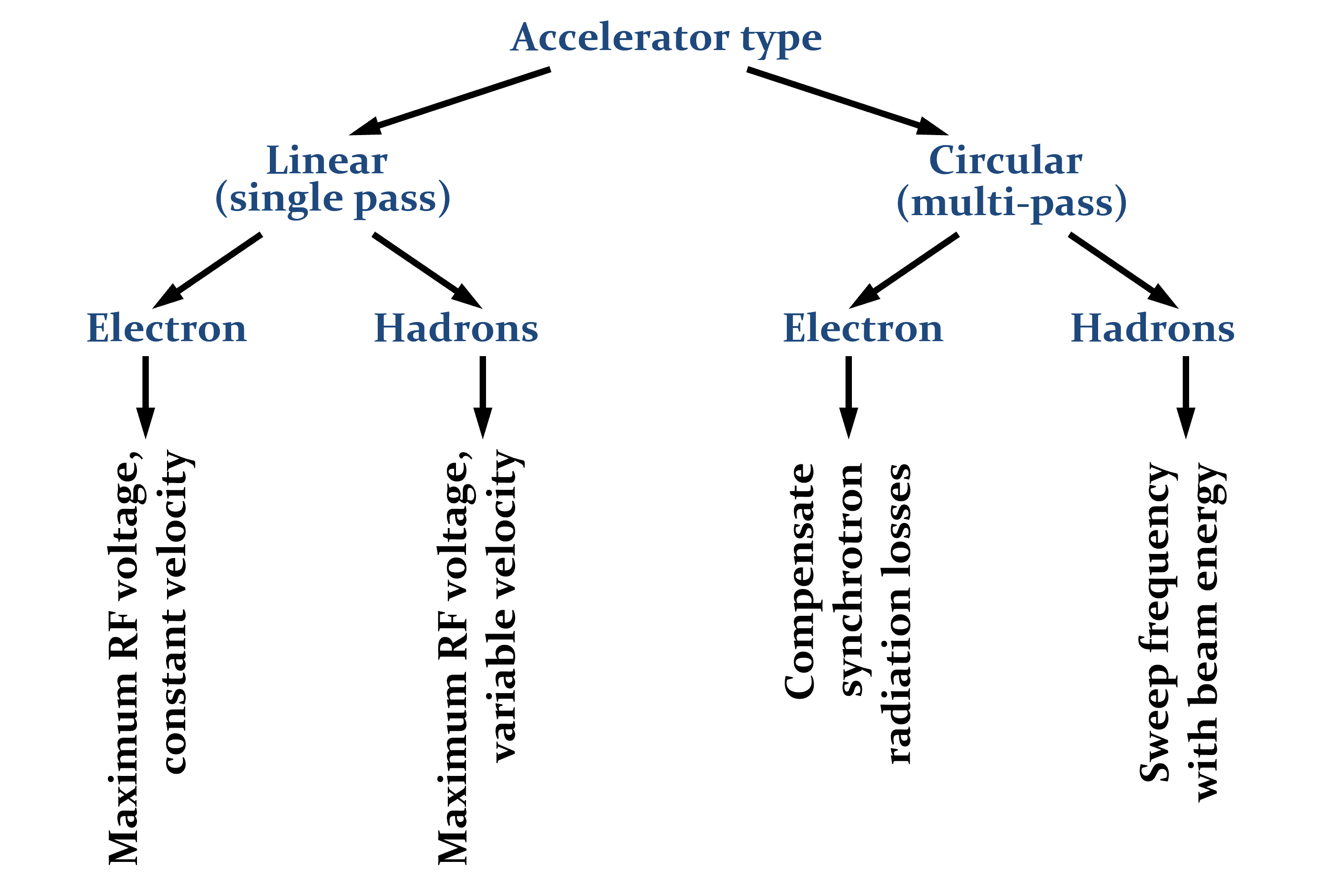}
	\caption{Main requirements for RF systems in different accelerator types.}
	\label{fig:RFSystemsHighEnergyAccelerators}
\end{figure}

The main requirements for the RF system majorly depend on the accelerator type as well as on the particle type, lepton or hadron, to be accelerated. Without being exhaustive, Fig.~\ref{fig:RFSystemsHighEnergyAccelerators} illustrates these constraints for the main accelerator types.
Since the particles only pass once through the RF structures of a linear accelerator (linac), the RF system is designed to deliver the maximum possible voltage in the shortest physical length. Highest gradients are therefore a main design criterion. Electrons become ultra-relativistic at moderate energy in the $\mathrm{MeV}$ range, and electron linacs are therefore operated at constant frequency, assuming a particle velocity, $v$ close to $c$, the speed of light in vacuum ($\beta = v/c \simeq 1$). In hadron linacs the RF frequency and type of accelerating structure must be adapted to the particle velocity. The RF structure type is therefore very different in the low or high energy part of the accelerator. In the latter case, the hadron velocity also becomes close to the speed of light and RF structures optimized for $\beta = 1$ are employed.

Most circular accelerators profit from the advantage that the particles pass through the accelerating stations multiple times, which allows to reach high energies with a moderate energy gain per turn. For the particular case of a synchrotron the orbit length is kept constant by increasing bending field and RF frequency synchronously. However, the periodicity introduces the additional constraint that the RF frequency, $f_\mathrm{RF}$ must be an integer harmonic of the revolution frequency, $f_\mathrm{rev}$~\footnote{Strictly speaking the RF voltage must only be repetitive with the revolution period.},
\begin{equation}
    f_\mathrm{RF} = n \cdot f_\mathrm{rev} \hspace{1em} \mathrm{with} \hspace{1em} n=1, 2, \ldots
    \label{eqn:rfFrequencyRevolutionFrequencyRelation} \, .
\end{equation}
For ultra-relativistic particles this can be conveniently achieved by an RF system at fixed frequency. In low- and medium-energy synchrotrons, in which the revolution frequency changes during acceleration, the RF system must also sweep its frequency during acceleration. This introduces important technical challenges. It is worth noting that rare exceptions exist to Eq.~(\ref{eqn:rfFrequencyRevolutionFrequencyRelation}). As only the RF voltage at the phase of the accelerated particles must be periodic, a pulsed RF system switched on and off every turn can be operated at virtually any fixed frequency to fulfill the periodicity condition~\cite{bib:boussard1989}.

\subsection{RF frequency}

The RF frequency is an important design parameter parameter of any accelerator which must be chosen carefully. As shown above it mainly depends on particle and accelerator type. However, the choice is not only guided by beam dynamics arguments, but also technical constraints which give preference to a specific RF frequency~\cite{bib:pirkl2005}. The final frequency choice is therefore often a compromise balancing the different requirements.

The advantages and disadvantages of a low RF frequency are summarized in Table~\ref{tab:whyChooseLowFrequency}.
\begin{table}[htb]
	\caption{Non-exhaustive list of advantages and disadvantages for a low RF frequency.}
	\label{tab:whyChooseLowFrequency}
	\centering\small
	\begin{tabular}{ll}
		\hline\hline \\[-0.5em]
		\textbf{Advantages} & \textbf{Disadvantages} \\[0.5em] \hline \\[-0.5em]
		
		\tabitem Large beam aperture & \tabitem Bulky cavities, size scales $\propto 1/f$, \\
		& \phantomtabitem volume $\propto 1/f^3$ \\[0.5em]
		
		\tabitem Long RF buckets, large acceptance & \tabitem Lossy material to downsize cavities \\ \\[0.5em]
		
		\tabitem Wide-band or wide range tunable & \tabitem Moderate or low acceleration gradient \\
		\phantomtabitem cavities possible \\[0.5em]

		\tabitem Power amplification and transmission & \tabitem Short particle bunches difficult \\
		\phantomtabitem straightforward & \phantomtabitem to generate \\[0.5em]
		
		\hline \hline

	\end{tabular}
\end{table}
For hadron synchrotrons in the low- and medium-energy range~(up to typically few tens of $\mathrm{GeV}$), RF frequencies below approximately $100\unsty{MHz}$ cannot be avoided as there is presently no technology to build wideband or wide-range tunable RF cavities in the frequency range beyond. Generally, RF frequencies below about $200\unsty{MHz}$ are preferred for most hadron linear accelerators, cyclotrons, as well as low- and medium energy hadron synchrotrons. The opposite Table~\ref{tab:whyChooseHighFrequency} summarizes main arguments in favor and against a high frequency for acceleration.
\begin{table}[htb]
	\caption{Non-exhaustive list of advantages and disadvantages for a high RF frequency.}
	\label{tab:whyChooseHighFrequency}
	\centering\small
	\begin{tabular}{ll}
		\hline\hline \\[-0.5em]
		\textbf{Advantages} & \textbf{Disadvantages} \\[0.5em] \hline \\[-0.5em]
		
		\tabitem Cavity size scales $\propto 1/f$, & \tabitem Available beam aperture \\
		\phantomtabitem volume $\propto 1/f^3$ & \phantomtabitem scales $\propto 1/f$ \\[0.5em]
		
		\tabitem Breakdown voltage increases & \tabitem No technology for wide-band or \\
		& \phantomtabitem tunable cavities \\[0.5em]
		
		\tabitem High gradient per unit length & \tabitem Power amplifiers more  difficult \\
		& \phantomtabitem to build \\[0.5em]
		
		\tabitem Particle bunches are short & \tabitem Power transmission losses \\[0.5em]
		
		\hline \hline

	\end{tabular}
\end{table}
RF frequencies beyond approximately $200\unsty{MHz}$ are thus mainly found in most linacs, in particular for leptons. Furthermore, a high RF frequency is the preferred choice for electron storage rings as well as high-energy hadron storage rings, where the frequency remains essentially constant during acceleration.

\begin{figure}[htb]
	\centering
	\includegraphics[width=0.7\linewidth]{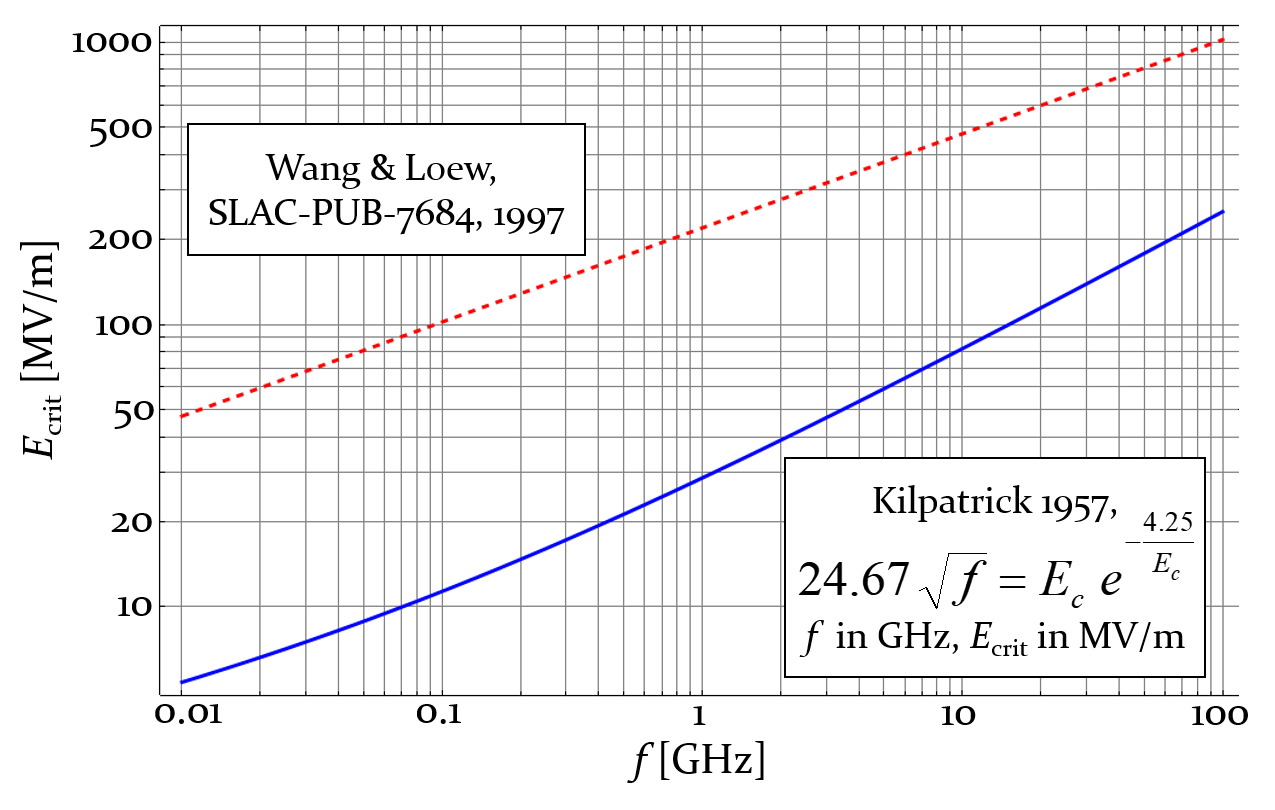}
	\caption{Maximum surface electric field versus frequency. The empirical criterion (blue) suggested by Kilpatrick in 1957~\cite{bib:kilpatrick1957} is compared to more recent measurements (red, dashed) which demonstrated significantly higher breakdown field~\cite{bib:wang1997}. Courtesy of E.~Jensen.}
	\label{fig:Kilpatrick}
\end{figure}

Not only the cavity size and volume shrink with increasing frequency, but the breakdown field limiting the maximum surface electric field in vacuum becomes more comfortable at higher frequencies as well~(Fig.~\ref{fig:Kilpatrick}).
Although physically not accurate, the Kilpatrick criterion has evolved to a reference unit describing the quality of an RF cavity in terms of breakdown voltage. Since its publication, much higher surface electric fields have been demonstrated, mainly due to better surface treatment. This improvement is commonly expressed as the factor with respect to the curve described by Kilpatrick. From Fig.~\ref{fig:Kilpatrick} it becomes clear that few $\mathrm{MV/m}$ at low frequencies are impossible to generate while even the $100\unsty{MV/m}$ regime becomes accessible in the $\mathrm{GHz}$ range.

In most cases the beam physics and engineering constraints only define a certain favorable frequency range for the RF system of an accelerator. This leaves some leeway for the exact frequency choice. In the past decades a set of common RF frequencies has emerged. Adhering to one of these frequencies assures that off-the shelf RF components are available from industry. Additionally, it allows the exchange of RF development and equipment amongst research laboratories using the same frequency. Table~\ref{tab:standardRfFrequency} summarizes a few standard RF frequencies.
\begin{table}[htb]
	\caption{Common RF frequencies in particle accelerators.}
	\label{tab:standardRfFrequency}
	\centering\small
	\begin{tabular}{ll}
		\hline\hline \\[-0.5em]
		\textbf{Accelerator} & \textbf{Frequency} \\[0.5em] \hline \\[-0.5em]
		Hadron accelerators and storage rings (RHIC, SPS) & $\simeq200\unsty{MHz}$ \\
        Electron storage rings (LEP, ESRF, Soleil) & $352\unsty{MHz}$ \\
        Electron storage rings (DORIS, PETRA, BESSY, SLS, etc.) & $499.6\ldots499.8\unsty{MHz}$ \\
        Superconducting electron linacs and FELs (X-FEL, ILC) &	$1300\unsty{MHz}$ \\
        Normal conducting electron linacs (SLAC) & $2856\unsty{MHz}$ \\
        High-gradient electron linac (CLIC) & $11.99 \unsty{GHz}$ \\
		\hline \hline
	\end{tabular}
\end{table}
While the RF systems of circular hadron accelerators are at the lower end of the frequency range, the frequency increases with the required gradient. To the reach gradients of the order of $100\unsty{MeV/m}$ for a compact linear collider (CLIC), an RF frequency above $10\unsty{GHz}$ is at present the only technically viable option.

\subsection{RF voltage}

\begin{figure}[htb]
	\centering
	\includegraphics[width=0.3\linewidth]{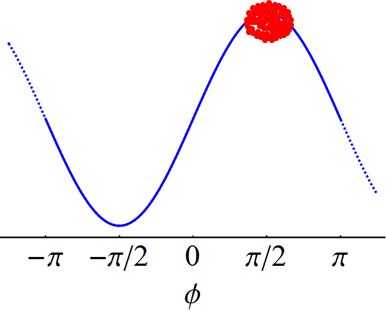} \hspace*{4em}
	\includegraphics[width=0.3\linewidth]{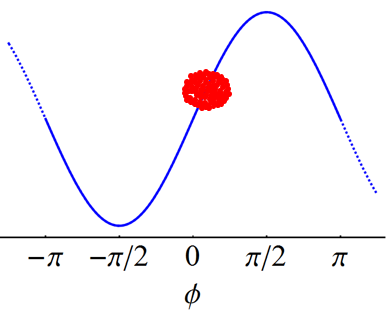}
	\caption{With the particles at the peak of the RF voltage (on-crest acceleration), the energy gain is maximized (left). Accelerating off-crest leaves margin for phase focusing (right).}
	\label{fig:RFVoltageOnOffCrestAcceleration}
\end{figure}

Next to the frequency the voltage is the second main design parameter for an RF system in a particle accelerator. At minimum the RF system must provide the required energy gain
\begin{equation*}
    qV = \Delta E \, ,
\end{equation*}
where $q$ is the particle charge and $V$ the accelerating voltage. During this so-called on-crest acceleration the particles are at the phase of the RF voltage maximum (Fig.~\ref{fig:RFVoltageOnOffCrestAcceleration}, left).
This kind of acceleration is employed in some linacs in the case of ultra-relativistic particles. Particles not located at the peak of the RF voltage, due to a phase error, do not receive the maximum energy gain and will lag behind after a few passages through the accelerating structures. This technique is hence not well suited for linacs at non-relativistic energies and circular accelerators.

More RF voltage needs to be provided to place the bunch of particles off-crest, at a so-called synchronous phase, $\phi_\mathrm{S}$, such that 
\begin{equation*}
    qV > \Delta E \hspace{1em} \rightarrow \hspace{1em} qV \sin \phi_\mathrm{S} = \Delta E \, .
\end{equation*}
This scheme ensures the phase focusing and keeps the particles together in bunches. For a particle with positive energy offset arriving before a particle with the reference energy, the RF voltage is lower than for the synchronous particle ~(Fig.~\ref{fig:RFVoltageOnOffCrestAcceleration}. right).  Hence, the energy gain  at the cavity passage is less than $qV \sin \phi_\mathrm{S}$. The particle with positive energy offset is therefore slowed down. The opposite applies to a particle arriving at the accelerating structure with too low energy. The stable phase is generally below $40^{\circ}$ for hadron synchrotrons and even smaller, in the range of $20^{\circ}$ for electron storage rings.

In a synchrotron or storage ring the RF system must firstly provide the energy gain per turn due to the changing field in the bending magnets. To keep the radius of the orbit constant, the centripetal force, $F_\mathrm{Z}$, must be equal to the Lorentz force, $F_\mathrm{L}$, of the dipole magnets:
\begin{equation*}
    F_\mathrm{Z} = F_\mathrm{L} \hspace{1em} \rightarrow \hspace{1em} \frac{p}{q} = \rho B \, ,
\end{equation*}
where $p$ and $q$ are momentum and charge of the particle; $\rho$ is the bending radius. A time dependent bending field, $dB/dt = \dot{B}$ will hence require a momentum change of 
\begin{equation*}
    \dot{p} = q \rho \dot{B} \, .
\end{equation*}
Discretizing the momentum change according to
\begin{equation}
    \dot{p} = \frac{ \Delta p }{ \Delta t } = \frac{m_0 c^2 \beta}{2 \pi R} \left( \beta \Delta \gamma + \gamma \Delta \beta \right) = \frac{\Delta E_\mathrm{turn}}{2 \pi R}
    \label{eqn:momentumRateVersusEnergyGainPerTurn}
\end{equation}
results in the synchronous energy gain per turn of
\begin{equation}
    \Delta E_\mathrm{turn} = 2 \pi q \rho R \dot{B}
\end{equation}
for a given ramp rate of the bending field. The total RF voltage installed in a synchrotron therefore scales with the size of the accelerator at fixed ramp rate, $\dot{B}$. Assuming that the ramp rate of the bending magnets is, in first order, independent of the circumference, the energy gain according to Eq.~(\ref{eqn:momentumRateVersusEnergyGainPerTurn}) required increases with the size. For example, the main accelerating system of the Proton Synchrotron (PS) at CERN (circumference: $2 \phi R = 2 \pi \cdot 100 \unsty{m}$) delivers a peak voltage of $200\unsty{kV}$, while the eleven times larger SPS is already equipped with an RF system capable of generating of the order of $10\unsty{MV}$. 

Secondly, the RF system must also compensate any energy loss of the particles, most prominently due to synchrotron radiation in lepton accelerators. The average energy loss per turn due to this effect becomes~\cite{bib:sands1970}
\begin{equation*}
    \Delta E_\mathrm{turn} = \frac{q^2}{3 \epsilon_0 (m_0 c^2)^4 } \frac{E^4}{\rho} \, ,
\end{equation*}
which can be expressed for electrons in the simplified equations between quantities
\begin{equation}
    \Delta E_\mathrm{turn} [\mathrm{keV}] = 88.5 \cdot \frac{ E^4 [\mathrm{GeV}]^4 }{ \rho [\mathrm{m}] } \hspace{1em} \mathrm{or} \hspace{1em} P_\mathrm{loss} [\mathrm{kW}] = 88.5 \cdot \frac{ E^4 [\mathrm{GeV}]^4 }{ \rho [\mathrm{m}] } \cdot I_\mathrm{B} [\mathrm{A}] \, .
\end{equation}
The particle rest mass is given by $m_0$, and $q$ is its charge; $\epsilon_0 \simeq 8.85 \cdot 10^{-12}\unsty{As/(Vm)}$ is the electric permittivity of free space. As synchrotron radiation losses are proportional to $1/m_0^4$, they are about thirteen orders of magnitude weaker for hadrons and only relevant in very high-energy synchrotrons with energies in the order of few $\mathrm{TeV}$.

\section{RF cavity}

The RF cavity is the most visible part of an RF installation. It is a oscillating system to achieve the high electric field gradients for the acceleration and manipulation charged particles in an accelerator by resonant excitation. It can be understood as coupler or adapter between the RF power from an amplifier and the beam. A wide range of technical implementations for RF cavities, from a simple pill-box cavity~\cite{bib:gerke1974,bib:wille2000} to the most sophisticated inter-digital~\cite{bib:ratzinger2005} or superconducting structures~\cite{bib:padamsee2008, bib:padamsee2013, bib:xiao2015}, can be found in particle accelerators. Depending on the application, a wide range of resonant frequencies, different geometries and technologies is employed.

Each resonant mode of an RF cavity can be described as a circuit of lumped elements.
\begin{figure}[htb]
	\centering
	\includegraphics[width=0.3\linewidth]{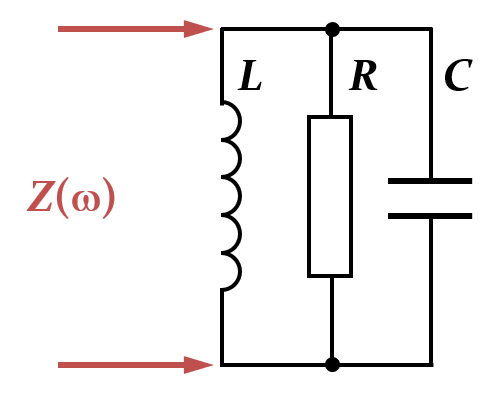}
	\caption{Lumped element representation of one resonance of an RF cavity.}
	\label{fig:LumpedElementCavityResonance}
\end{figure}
The impedance of this parallel circuit of resistance, $R$, inductance, $L$ and capacitance, $C$ is written as
\begin{equation}
    \frac{1}{Z(\omega)} = \frac{1}{R} + \frac{1}{i \omega L} + i \omega C \, ,
    \label{eqn:lumpedElementResonanceImpedance}
\end{equation}
which resonates at the frequency where the impedance of inductance and capacitance compensate, 
\begin{equation*}
    \omega_0 = \frac{1}{\sqrt{LC}} \, .
\end{equation*}
The quality factor, $Q$ of a resonant oscillator is proportional to the ratio of stored energy and the energy lost during an oscillation cycle
\begin{equation}
    Q = \omega_0 \frac{\text{Stored energy}}{\text{Average power loss}} = \frac{\omega_0 E}{P} \, .
\end{equation}
The energy stored in the resonant mode is distributed between both reactive elements. In the limiting case the energy
\begin{equation}
    E = \frac{1}{2} C V^2 = \frac{1}{2} L I^2
\end{equation}
is either be stored in capacity or inductance alone, where $V$ is the maximum voltage and $I$ the maximum current, respectively. In terms of the lumped elements, the quality factor can hence be expressed in the form
\begin{equation}
    Q = \omega_0 RC = \frac{R}{\omega_0 L} \, .
    \label{eqn:cavityQualityFactorRelations}
\end{equation}

Replacing the reactive elements in Eq.~(\ref{eqn:lumpedElementResonanceImpedance}) by quality factor and resonance frequency, the resonator impedance becomes
\begin{equation}
    Z(\omega) = \frac{R}{1+iQ \left(\cfrac{\omega^2 - \omega_0^2}{\omega \omega_0} \right)} \simeq  \frac{R}{2iQ \frac{\Delta \omega}{\omega_0}} \, ,
    \label{eqn:resonanceImpedanceInResonanceFrequencyAndGeometryFactor}
\end{equation}
where the approximation applies at frequencies close to the resonance, $\Delta \omega \ll \omega_0$.

Comparing Eq.~(\ref{eqn:lumpedElementResonanceImpedance}) with Eq.~(\ref{eqn:resonanceImpedanceInResonanceFrequencyAndGeometryFactor}) it becomes clear that the resonance can be equivalently described by the lumped elements $R$, $L$, $C$, as well as by the parameters $R$, $R/Q$ and $\omega_0$ or any other set of three parameters.

While the lumped element representation for a resonant mode of a cavity is convenient for circuit simulations, it is not adapted to the design of RF cavities. In a particle accelerator the resonance frequency, $\omega_0$ of a cavity is normally given by the beam parameters, for example a specific harmonic of the revolution frequency in a synchrotron or storage ring. Neglecting energy transfer into the beam, the shunt impedance $R$ relates the achievable gap voltage with the available RF power. The third parameter most relevant to the cavity designer is the ratio of shunt impedance and quality factor, the so-called ``R-upon-Q'', $R/Q$. This parameter is only defined by the cavity geometry and is thus a main criterion to optimize a cavity design.

\subsection{Passage of a particle charge through an RF cavity}
\label{sec:PassageParticleChargeThroughCavity}

\begin{figure}[htb]
	\centering
	\includegraphics[width=0.28\linewidth]{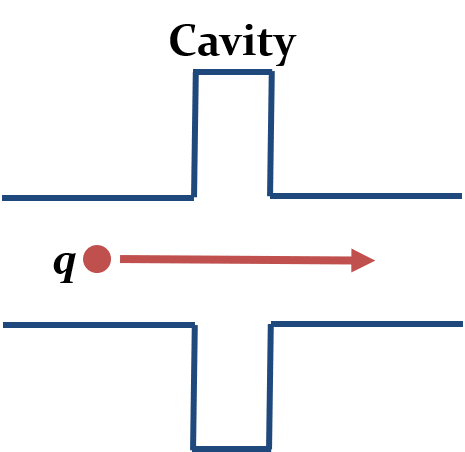} \hspace*{3em}
	\includegraphics[width=0.4\linewidth]{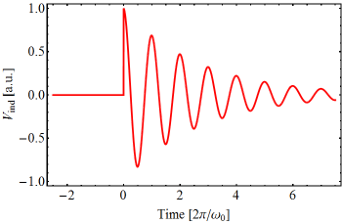}
	\caption{Charge passing through an RF cavity (left) and resulting beam induced voltage (right).}
	\label{fig:CavitySketchesWithChargePassingThrough}
\end{figure}

When a charge, $q$ travels through a cavity at the speed of light~(Fig.~\ref{fig:CavitySketchesWithChargePassingThrough}),
it only interacts with the electric field of the cavity in first order approximation. In the instant of the passage the resonator behaves like a capacity, $C$ which is charged since
\begin{equation*}
    q = V_\mathrm{ind} C \, .
\end{equation*}
The beam induced voltage at the cavity gap is $V_\mathrm{ind}$. This very simplified model assumes also that the charge is not slowed down due to the energy loss.

Using Eq.~(\ref{eqn:cavityQualityFactorRelations}) the capacity $C$ of the resonator can be replaced by $R/Q$ since
\begin{equation*}
    \frac{1}{C} = \left( \frac{R}{Q} \right) \omega_0
\end{equation*}
and the beam induced voltage can be written as
\begin{equation}
    V_\mathrm{ind} = \frac{q}{c} \propto \frac{R}{Q} \, .
    \label{eqn:cavityResonatorBeamInducedVoltage}
\end{equation}
Equation~(\ref{eqn:cavityResonatorBeamInducedVoltage}) illustrates that the beam loading, which is the ability of the beam to deposit energy in a cavity is directly proportional to the geometry factor $R/Q$. In high current accelerators, a low $R/Q$ is preferred. Superconducting cavities with high quality factor are employed to achieve high shunt impedance at moderate $R/Q$.

\subsection{Cavity geometries}

Depending on the required frequency RF cavities for acceleration may have very different geometries. However, as will be shown by the following examples, most of them can be reduced to the same basic resonator types.

\subsubsection{Line resonators}
\label{Sec:LineResonators}

As the cavity size scales with the RF wavelength $\lambda \propto 1/f$ and their volume even with $1/f^3$, the size cavities in the low frequency range quickly becomes unmanageable. Below $10\unsty{MHz}$ the RF wavelength is in excess of $30\unsty{m}$. This would ask for huge cavities, too large for the space available in the straight sections in low and medium energy synchrotrons.

\begin{figure}[htb]
	\centering
	\includegraphics[width=0.75\linewidth]{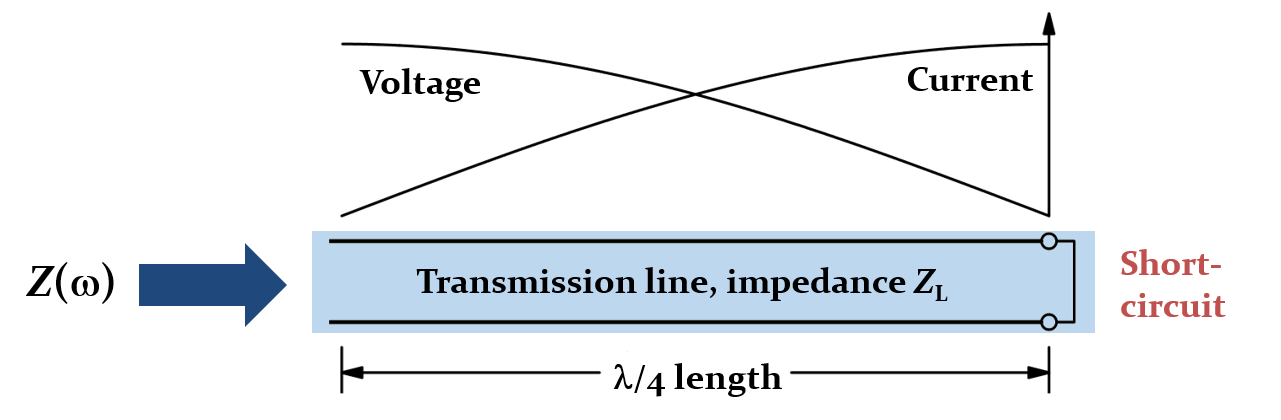}
	\caption{Voltage and current in a line resonator. The length is chosen to be a quarter of the wavelength}.
	\label{fig:LineResonatorSketch}
\end{figure}

A very common resonator type to overcome this problem is the so-called line resonator~(Fig.~\ref{fig:LineResonatorSketch}).
A transmission line of the characteristic impedance $Z_\mathrm{L}$ is short-circuited on one side, provoking ideally zero voltage and a large current. The line of the length $\lambda/4$ is left open on the other side such that no current can flow, but voltage develops.

\begin{figure}[htb]
	\centering
	\includegraphics[width=0.75\linewidth]{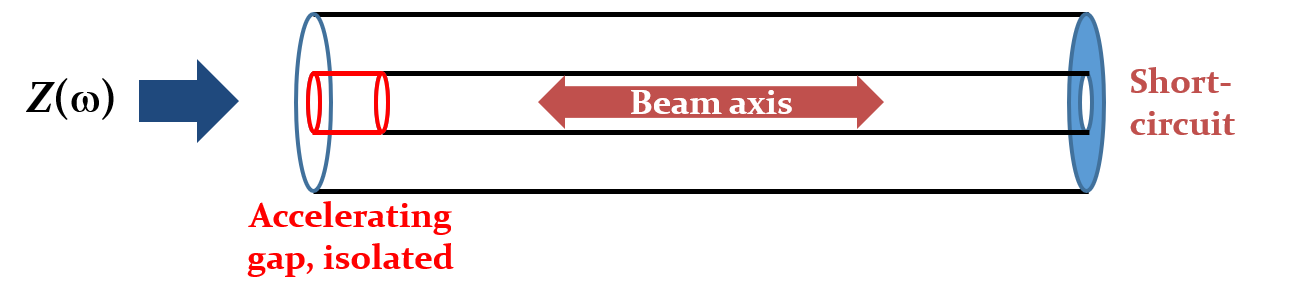}
	\caption{Coaxial design of a $\lambda/4$-resonator. The inner conductor conveniently also acts a beam pipe.}.
	\label{fig:CoaxialLambaQuarterResonatorSketch}
\end{figure}

Such a $\lambda/4$ resonator is a very common cavity type in particle accelerators for frequencies in the range below $100\unsty{MHz}$. Built as a coaxial system~(Fig.~\ref{fig:CoaxialLambaQuarterResonatorSketch}), the inner conductor simply serves as a beam pipe.
The particles hence remain shielded inside the inner conductor until they reach the high voltage side of the $\lambda/4$-resonator. The longitudinal electric field is generated at the gap between the end of the inner conductor and the end plate. However, at low frequencies in the 10 MHz range and below, the longitudinal dimensions of a $\lambda/4$-resonator are still large. One can therefore introduce additional capacity or inductance to lower the resonance frequency while keeping the mechanical cavity dimensions reasonable.

\begin{figure}[htb]
    \centering
	\includegraphics[width=0.48\linewidth]{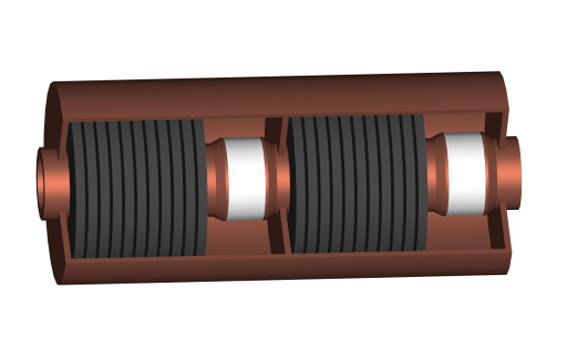} \hspace*{2em}
	\includegraphics[width=0.36\linewidth]{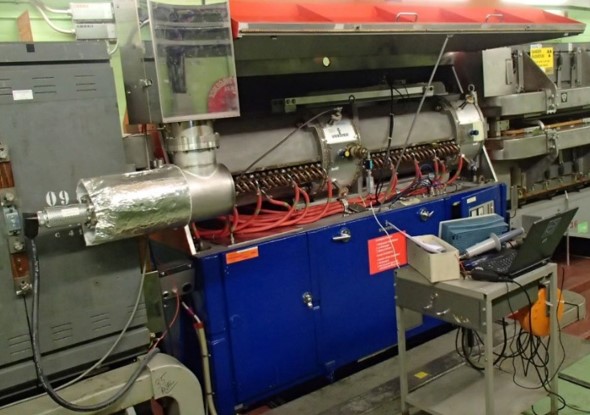}
	\caption{PS $10\unsty{MHz}$ main accelerating cavity as an example for an inductively shortened coaxial resonator. Two resonators are placed in series to keep the voltage across each gap in the $10\unsty{kV}$ range. The length of the entire unit is only $3\unsty{m}$.}
	\label{fig:ExamplePS10MHzCavity}
\end{figure}

An example of an inductively shortened cavity is the ferrite-loaded cavity~\cite{bib:gardner2000,bib:klingbeil2010} of the CERN PS~(Fig.~\ref{fig:ExamplePS10MHzCavity}).
Adding ferrite material with high permeability results in large line inductance per unit length, raising the characteristic impedance according to
\begin{equation*}
    Z_\mathrm{L} = \frac{1}{2 \pi} \sqrt{ \frac{ \mu_0 }{ \epsilon_0 } \frac{ \mu_\mathrm{r}}{ \epsilon_\mathrm{r} } } \ln \frac{D}{d} \simeq 60\unsty{\Omega} \sqrt{ \frac{ \mu_\mathrm{r}}{ \epsilon_\mathrm{r} } } \ln \frac{D}{d}
\end{equation*}
while lowering the propagation velocity to
\begin{equation*}
    v = \frac{c} { \sqrt{ \mu_\mathrm{r} \epsilon_\mathrm{r}  } } \, .
\end{equation*}
This shortens the cavity such that a quarter wavelength of more than $25\unsty{m}$ can fit into a compact cavity of physical length below $1.5\unsty{m}$. Even smaller cavities with larger voltage capability or larger bandwidth can be achieved with high-permeability magnetic alloy material like, for example, Finemet~\cite{bib:mori1998}.

\begin{figure}[htb]
    \centering
	\includegraphics[width=0.27\linewidth]{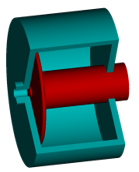} \hspace*{4em}
	\includegraphics[width=0.27\linewidth]{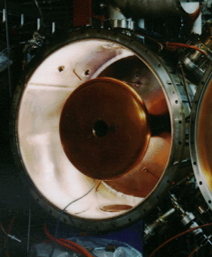}
	\caption{Capacitively shortened $53\unsty{MHz}$ cavity of the National Synchrotron Light Source (NSLS) at the Brookhaven National Laboratory (BNL)~\cite{bib:hanna1993}. Together with the end cap, the large plate at the inner conductor shortens the resonator.}
	\label{fig:ExampleNSLSCavity}
\end{figure}

Adding magnetic material like ferrite or magnetic alloy nonetheless introduces losses which lower the shunt impedance and increase the RF power requirements. An alternative technique to reduce the resonator frequency is capacitive loading. The capacity must be located in a region of large electric field and is hence in the region of the accelerating gap. Figure~\ref{fig:ExampleNSLSCavity} shows typical example of a capacitively shortened coaxial cavity. Since no lossy material is required in this case, larger shunt impedances are reached.

\subsubsection{Cavity resonators}

\begin{figure}[htb]
    \centering
	\includegraphics[height=0.21\linewidth]{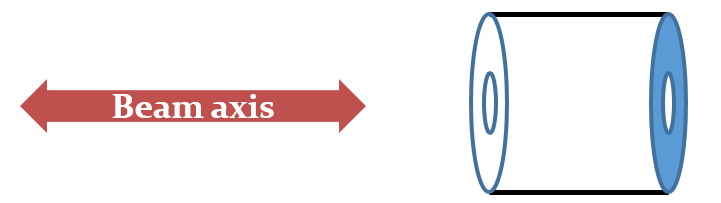} \hspace*{1em}
	\includegraphics[height=0.21\linewidth]{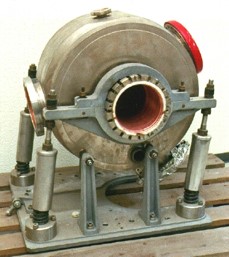}
	\caption{Basic pill box cavity resonator (left) and the example of the $500\unsty{MHz}$ accelerating cavity of the DORIS electron-positron storage ring.}
	\label{fig:ExamplePillBoxCavity}
\end{figure}

The main losses in conventional coaxial line resonators are caused by the RF current flow on the surface of the inner conductor. When moving to higher frequencies such a conductor is not needed. As illustrated in Fig.~\ref{fig:ExamplePillBoxCavity}, the cavity then becomes a simple cylinder, also known as pill-box cavity, with holes in both end plates to connect to the beam pipe.
Such simple pill box cavities have indeed been operated in the first generation of electron storage rings and synchrotron radiation light sources like the DORIS~\cite{bib:wille2000}, mostly at a resonance frequency around $500\unsty{MHz}$~(Fig.~\ref{fig:ExamplePillBoxCavity}, right).

To calculate the resonance frequency, the cavity can be considered as a section of circular wave guide. The transverse magnetic (TM) cut-off frequency of a cylindrical wave guide is given by the condition that the longitudinal electric field, $E_\mathrm{z}$ must vanish on the surface of the conducting cylinder. Transverse magnetic means that there is no longitudinal magnetic field, but only longitudinal electric field. This is equivalent to the condition
\begin{equation*}
    J_0 \left( k_\mathrm{c} \frac{D}{2} \right) = 0 \hspace{1em} \mathrm{with} \hspace{1em} k_\mathrm{c} = \frac{2 \pi}{\lambda_\mathrm{c}} \, ,
\end{equation*}
where $J_0(\chi)$ is the Bessel function of first kind, $D$ the diameter of the wave guide and $\lambda_\mathrm{c}$ its fundamental cut-off wavelength. With the zero crossing of the Bessel function being at $\chi_\mathrm{01} \simeq 2.40483$, the wavelength at cut-off is hence given by
\begin{equation}
    \lambda_\mathrm{c} = \frac{\pi D}{\chi_{01}} \, .
    \label{eqn:CutOffFrequencyCircularWaveguide}
\end{equation}
Closing the circular wave guide with end plates perpendicular to the wave guide axis creates a cavity of length, $l$. This extra boundary condition allows additional resonant modes. In combination with Eq.~(\ref{eqn:CutOffFrequencyCircularWaveguide}) it results in the resonance frequencies~\cite{bib:jensen2010,bib:jensen2013}
\begin{equation}
    \omega_{01n} = c \cdot \sqrt{ \left( \frac{2 \chi_{01}}{D} \right)^2 + \left( \frac{\pi n}{l} \right)^2 } \, \hspace{1em} \mathrm{with} \hspace{1em} n = 0, 1, 2, \ldots
\end{equation}
As the Bessel functions have multiple zero crossings, $\chi_{pq}$ the number of cavity resonances due to the cavity diameter $D$ is infinite.

The resonance frequency of the lowest order transverse magnetic (TM) mode, $\omega_{010}$ is independent of the cavity length, $l$. Only for a short pillbox cavity ($ l < 1.01538\,D$) this mode is the fundamental one. For a longer cavity the resonance frequency of the fundamental mode becomes a TE mode, which has a resonance frequency below $\omega_{010}$. In this case the frequency of the fundamental mode decreases with increasing cavity length. Only the TM mode is suitable for acceleration as its longitudinal electric field is largest on axis, where the beam passes through the cavity.

The presence of further resonant modes at higher frequencies, the so-called higher-order modes~(HOM), are an important issue in cavities installed in particle accelerators. In rings the longitudinal beam spectrum mainly consists of lines at integer multiples of the revolution frequency. Whenever one of these lines overlaps with a frequency of an HOM, beam power is unintentionally coupled back into the cavity. This does not only cause energy loss and heating in the cavity, but also may drive a high-intensity beam unstable.

More evolved cavity shapes have therefore been designed to optimize for the specific application in high-intensity accelerators. In the Large Hadron Collider (LHC) at CERN the average energy gain per turn is with only about $0.5\unsty{MV/turn}$ very moderate and the total RF voltage of up to $16\unsty{MV}$ per turn is only required to keep the bunches short in collision. However, as the average beam current of more than $0.5\unsty{A}$ is very large (nominal bunch intensity of $1.3 \cdot 10^{13}\mathrm{p/b}$ at $25\unsty{ns}$ spacing) and, due to the low revolution frequency, the longitudinal beam spectrum contains spectral lines every $11\unsty{kHz}$, beam loading and beam power coupling to HOMs are major issues.

\begin{figure}[htb]
    \centering
	\includegraphics[width=0.3\linewidth]{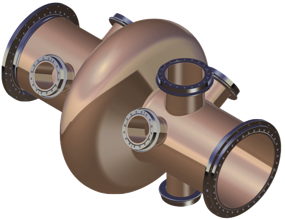} \hspace*{4em}
	\includegraphics[width=0.48\linewidth]{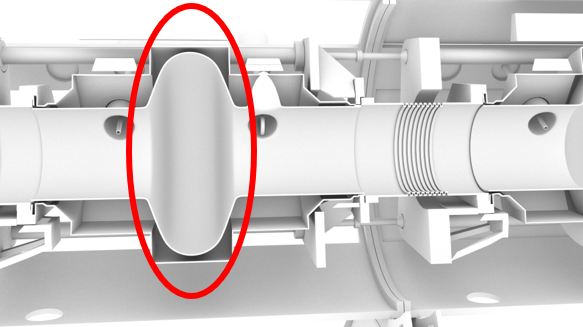}
	\caption{Three-dimensional drawing of a single cell LHC cavity, resonating at about $400\unsty{MHz}$ (left), as well as a cut through the installed cavity outlining the shape of the resonator (right)~\cite{bib:boussard1999}.}
	\label{fig:LHCCavityExample}
\end{figure}

A bell-shaped cavity design has been adopted~(Fig.~\ref{fig:LHCCavityExample}) to reduce beam loading at the fundamental resonance at about~$400.8\unsty{MHz}$.
Thanks to this geometry the $R/Q$ is as low as~$44\unsty{\Omega}$. To achieve a high quality factor to keep high shunt impedance, superconducting cavities were the only choice for the LHC. To reduce beam induced power at HOMs inside the resonator, single cell cavities connected by wide beam pipes were chosen~\cite{bib:boussard1993}. These beam pipes have a low wave guide cut-off frequency allowing the HOMs to propagate from the cavity into the beam pipes where special couplers are installed to damp them.  

\subsection{Cavity coupling}

To generate the electric field for acceleration in a cavity resonator, RF power from a source, an RF amplifier must be coupled into the cavity. There are several possibilities to excite the resonator which becomes also clear from the lumped element model of the resonance~(Fig.~\ref{fig:LumpedElementCavityResonance}). The coupling can be performed inductively through the magnetic field in the cavity, capacitively through the electric field near the gap, or as a combination of both. Technically this is achieved with antennas in the form of inductive loops or capacitive elements inserted into cavity. A cavity can moreover be excited by a wave guide attached to it~\cite{bib:alesini2010,bib:haebel1996}.

\subsection{Inductive coupling}

\begin{figure}[htb]
    \centering
	\includegraphics[height=0.25\linewidth]{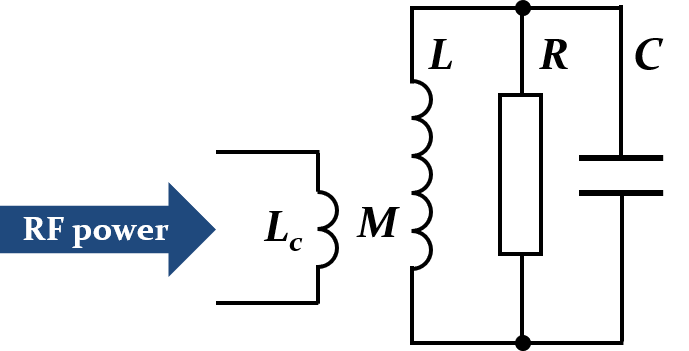} \hspace*{4em}
	\includegraphics[height=0.25\linewidth]{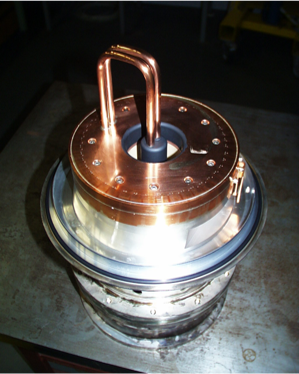}
	\caption{Equivalent schematics of inductive cavity coupling (left). The right photo shows the water-cooled loop to couple $1\unsty{MW}$ of power into the main accelerating cavities of the $50\unsty{MHz}$ cavities in the $590\unsty{MeV}$ cyclotron. Photo courtesy of L.~Stingelin.}
	\label{fig:CavityCouplingSchematicsInductive}
\end{figure}

The inductive or magnetic coupling is most common for cavities in the frequency range of up to a few $100\unsty{MHz}$. A loop is placed in a region of large magnetic field in a cavity. For the example of the line resonator~(Sec.~\ref{Sec:LineResonators}) this is near the short circuit side. Figure~\ref{fig:CavityCouplingSchematicsInductive} shows the equivalent circuit model~(left) together with an example of the high-power coupling loop to transfer about $1\unsty{MW}$ continuous wave at $50\unsty{MHz}$ into the accelerating cavities of a cyclotron.
The loop itself is characterized by its self-inductance, $L_\mathrm{c}$. Together with inductance of the resonator, it shapes a transformer with the mutual inductance $M$.

Loop coupling has several advantages. The coupling $M$ can be easily adjusted by simply rotating the loop in the magnetic field to conveniently vary the effective surface. As the gap region with maximum electric field is different from the regions of high magnetic field, the loop is installed far away from the gap and hence introduces only little perturbation. Technically, high power levels can be reached easily by flowing cooling water through the loop and, thanks to the low impedance, the involved voltages remain moderate.

\subsection{Capacitive coupling}

\begin{figure}[htb]
    \centering
	\includegraphics[height=0.25\linewidth]{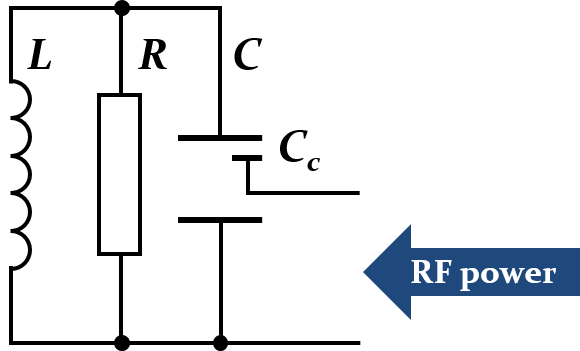} \hspace*{4em}
	\includegraphics[height=0.25\linewidth]{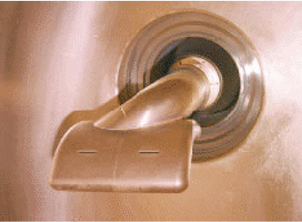}
	\caption{Inductive (left) and capacitive (right) coupling of RF power into a resonator~\cite{bib:garoby1997}.}
	\label{fig:CavityCouplingSchematicsCapacitive}
\end{figure}

Less common is capacitive or electric coupling directly to the cavity, which connects to the resonator in the same way as the beam (Fig.~\ref{fig:CavityCouplingSchematicsCapacitive}, left). 
The $40\unsty{MHz}$ cavities in the CERN PS are coupled capacitively through an arm~(Fig.~\ref{fig:CavityCouplingSchematicsCapacitive}, right) which is brought close to the cavity gap. The position of the arm is aligned to achieve the intended coupling ratio. The main advantage of capacitive coupling becomes apparent in combination with tube-based power amplifiers. The tube needs a large anode voltage in its output circuit. Since the capacitive arm is isolated from the cavity, it can be driven with a combination of RF voltage and anode voltage which removes the need for a technically difficult de-coupling capacitor.

\begin{figure}[htb]
    \centering
	\includegraphics[height=0.35\linewidth]{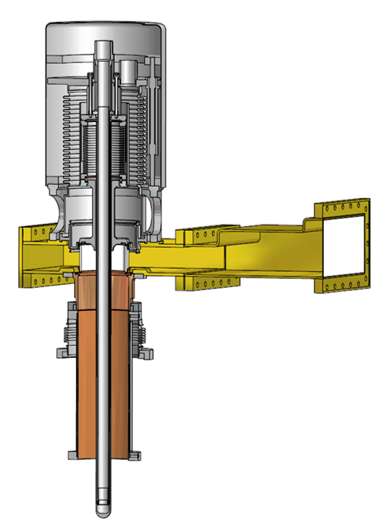}
	\caption{Electric coupling of RF power into a resonator through an antenna~\cite{bib:kindermann1999}.}
	\label{fig:CavityCouplingSchematicsCombined}
\end{figure}

As an example for electric coupling through an antenna, the power coupler of the main accelerating cavities of the LHC at $400.8\unsty{MHz}$ is illustrated in Fig.~\ref{fig:CavityCouplingSchematicsCombined}.
It allows to continuously transfer about $300\unsty{kW}$ into each cavity. To cope with the different conditions during injection, acceleration and collision in the LHC, it features a mechanically changing coupling ratio. During the injection process, where beam loading should be minimized, the antenna is plunged deeper into the cavity structure, linking it more tightly to the power amplifiers. In collision, when the maximum RF voltage is required, the antenna is partly removed from the cavity to reduce the coupling and increase the effective shunt impedance.

\section{RF power amplifier}

The RF amplifier, sometimes also called power plant, is connected to the cavity resonator and supplies the energy for the acceleration of the beam. Two major different technologies co-exist to amplify an RF signal. Firstly, vacuum tube amplifiers play an important role in accelerators. As the physical size of vacuum electronics grows with the output power, gridded tubes like triodes and tetrodes may not be considered small with respect to the RF wavelength anymore. The fact that a device has a size comparable to or is even larger than the wavelength is actually employed in travelling wave tubes, klystrons and many other types of tubes. In-between, the inductive output tube (IOT) unites the moderate power input circuitry of a triode with the high-power output system of a klystron. Secondly, solid state amplifiers have become a mature competitor, even at highest power levels. While transistors themselves as amplifying elements are much smaller than tubes, the auxiliary circuitry requires significant space. The achievable power per transistor remains still significantly below the power from a single tube device. Hence the outputs of many transistors, typically hundreds to thousands, are combined to achieve the desired power levels. The gain per amplification stage moreover defines how many stages must be chained to obtain the required overall gain. Travelling wave tubes may achieve significantly higher gains than gridded tubes or transistors. This may simplify an installation.

\begin{figure}[htb]
    \centering
	\includegraphics[height=0.35\linewidth]{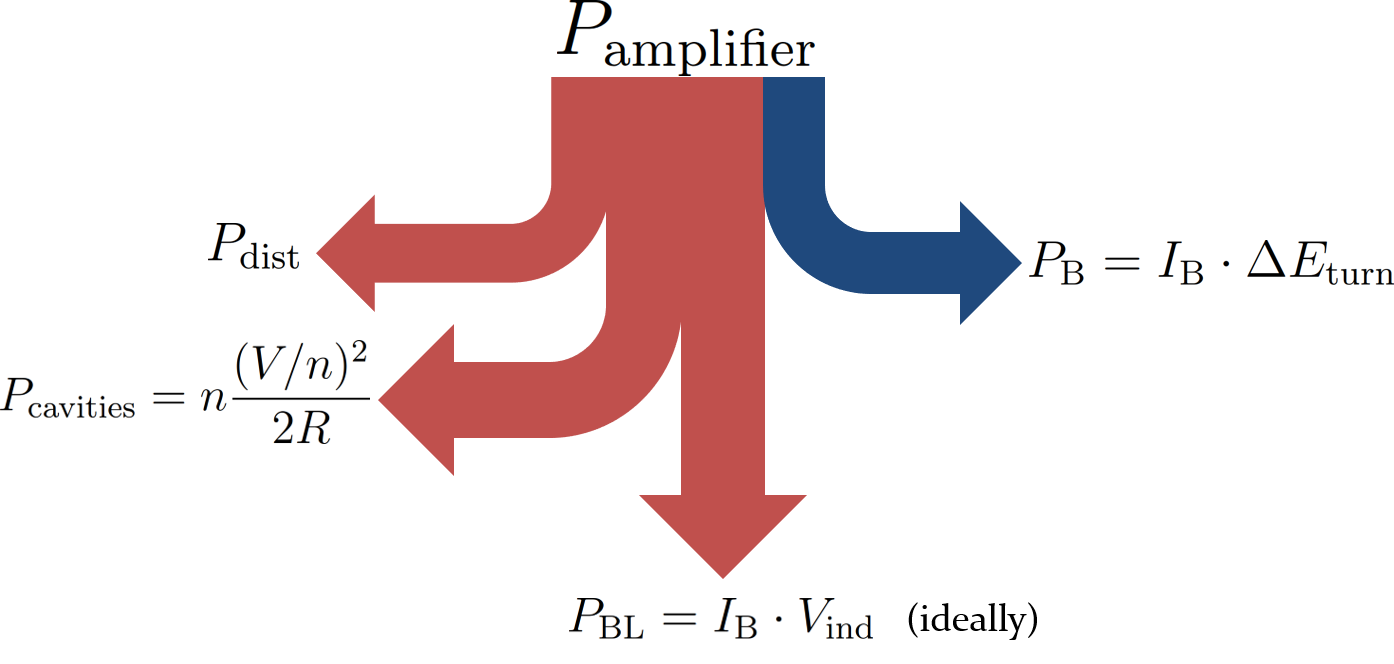}
	\caption{Power balance of an RF amplifier driving an RF cavity.}
	\label{fig:RFAmplifierPowerDistribution}
\end{figure}

Ideally, the available output power of the amplifier, $P_\mathrm{amplifier}$ would be entirely transferred to the beam. The power balance in a real RF system as illustrated in Fig.~\ref{fig:RFAmplifierPowerDistribution} is very different.
Next to the power delivered to accelerate a beam current, $I_\mathrm{B}$ by an energy, $\Delta E$
\begin{equation*}
    P_\mathrm{B} = I_\mathrm{B} \cdot \Delta E \, ,
\end{equation*}
some power, $P_\mathrm{dist}$ may already be lost in the feeder line between the amplifier and the cavity. Furthermore, the losses in the cavities
\begin{equation*}
    P_\mathrm{cavities} = n \frac{(V/n)^2}{2R} \, ,
\end{equation*}
where $n$ is the number of RF stations, must be compensated just to produce the total accelerating voltage,~$V$. This equation suggests that distributing the RF voltage over several cavities is beneficial for the total RF power which only grows proportionally to their number. Large particle accelerators are therefore equipped with an ensemble of RF stations.

Finally, the energy does not only flow from the amplifier through the cavity to the beam, but the beam also delivers energy back to the RF system, as introduced in Sec.~\ref{sec:PassageParticleChargeThroughCavity}. Ideally compensating the beam induced voltage would require a power of
\begin{equation*}
    P_\mathrm{BL} = I_\mathrm{B} \cdot V_\mathrm{ind} \, ,
\end{equation*}
Fully removing the beam induced voltage contribution in high current accelerators would require excessive RF power and partial compensation schemes to mitigate beam loading effect are put in place~\cite{bib:boussard1991,bib:baudrenghien2007,bib:mastoridis2017}.

Conceptually a power amplifier ideally just multiplies the input signal by a constant factor
\begin{equation}
    P_\mathrm{out} = g \cdot P_\mathrm{in} \hspace{1em} \mathrm{or} \hspace{1em} V_\mathrm{out} = \sqrt{g} \cdot V_\mathrm{in} \, ,
\end{equation}
which does not pay tribute the complexity of a real device.
For an ideal amplifier, however, the power amplification factor $g$ would be the only parameter. Such an ideal amplifier would have a large bandwidth to amplify all frequencies equally well, no saturation and infinite RF power with zero delay. It would furthermore not add any noise to the amplified input signal, be unconditionally stable under all load conditions and resist reverse power which may be delivered back from the cavity to the amplifier. Radiation hardness may be  another requirement in the particle accelerator environment.

\subsection{Diodes, triodes and tetrodes}

\begin{figure}[htb]
    \centering
	\includegraphics[height=0.25\linewidth]{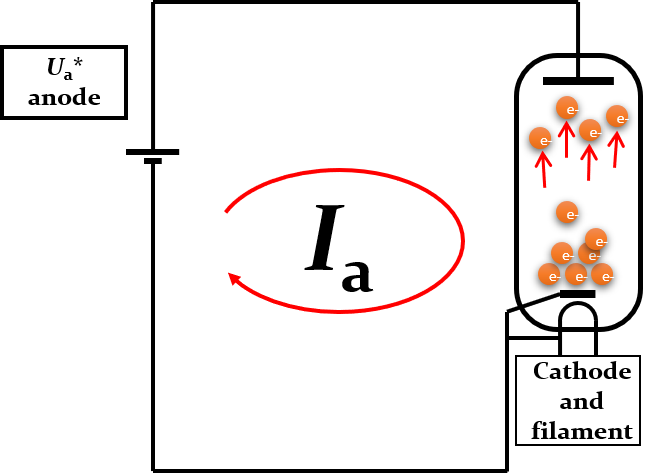} \hspace*{4em}
	\includegraphics[height=0.25\linewidth]{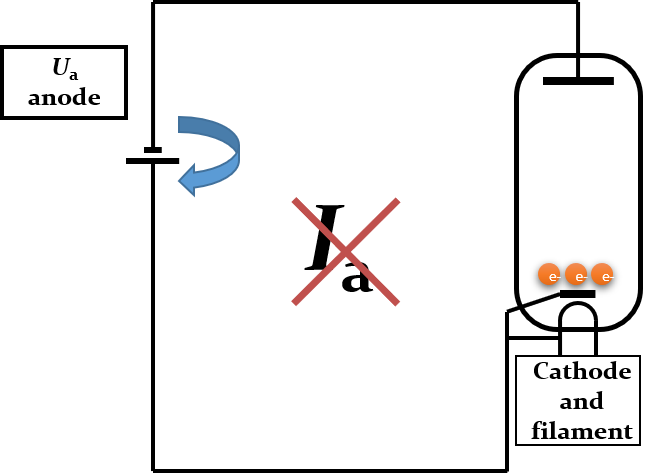}
	\caption{Vacuum diode in conducting (left) and blocking state (right)~\cite{bib:montesinos2018}. Images courtesy of E. Montesinos.}
	\label{fig:VacuumDiodeConductingBlocking}
\end{figure}

The principle of operation of valves like diodes, triodes or tetrodes is based on an electric current generated by a flow of electrons in vacuum. The electrons are released by thermionic emission. The so-called cathode electrode is heated by a filament and produces an electron cloud. Coated metals, carbides or borides are often used as materials due to their low energy required to liberate these electrons. Applying a positive voltage to a second electrode, the anode, generates an electric field which attracts the electrons and causes a current flow (Fig.~\ref{fig:VacuumDiodeConductingBlocking}, left).
With a negative voltage at the anode the electrons are repelled preventing any current.

\begin{figure}[htb]
    \centering
	\includegraphics[height=0.25\linewidth]{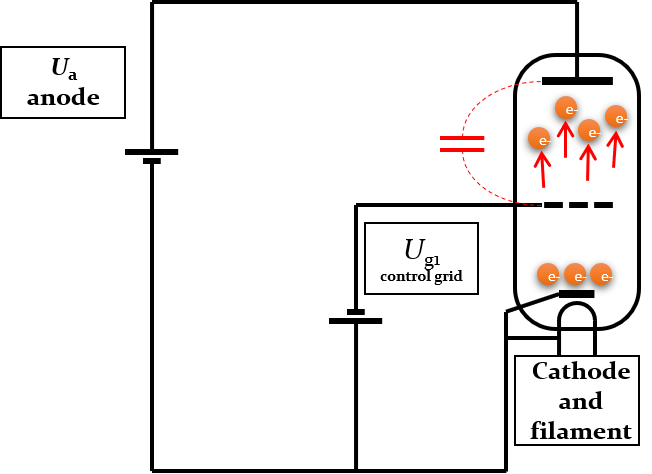} \hspace*{4em}
	\includegraphics[height=0.25\linewidth]{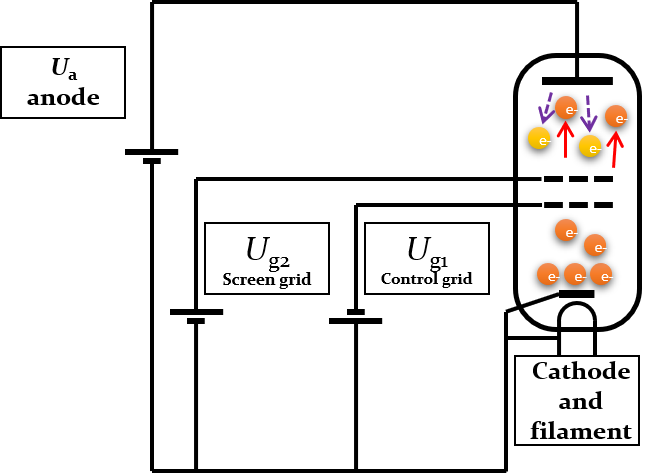}
	\caption{Vacuum tube with three electrodes (triode, left) and with additional shielding grid (tetrode, right)~\cite{bib:montesinos2018}. Images courtesy of E. Montesinos.}
	\label{fig:VacuumTriodeTetrode}
\end{figure}

Introducing a grid, $g_1$ between cathode and anode as illustrated in Fig.~\ref{fig:VacuumTriodeTetrode} (left) allows to control the current flow between cathode and anode. 
The control grid does not attract any electrons which could cause current flow as long as it is kept at a negative bias voltage with respect to the cathode. Controlling the anode current, $I_\mathrm{a}$ through the grid voltage, $U_\mathrm{g1}$ can be expressed as the transconductance
\begin{equation*}
    g_\mathrm{m} = \frac{ \Delta I_\mathrm{a} }{ \Delta U_\mathrm{g1} } \, ,
\end{equation*}
which describes the gain of this triode in an RF power amplifier. For this type of tube the gain is limited by the parasitic capacity between the anode and the control grid. The amplified signal at the anode couples back to the grid which causes the amplifier to excite itself to uncontrolled oscillations.

Damping of these oscillations is achieved by reducing the capacitive coupling between anode and control grid (Fig.~\ref{fig:VacuumTriodeTetrode}, right). A second grid, the so-called screen grid, $g_2$, is introduced near the control grid. When biased at a positive voltage with respect to cathode and control grid, but still well below the anode voltage, it removes the capacitive coupling between control grid and anode.

Tetrodes are the main work horse of high-power RF amplifiers. They combine large gain, thanks to the additional pre-acceleration of the electrons, with good stability. Their performance is mainly limited by secondary electrons which are released at the anode due to the impacting primary, higher-energy electrons from the cathode and pre-accelerated by the positive voltage at the screen grid. The secondary electrons may cause a current flow through the screen grid, which can ultimately damage its fine mesh structure. The secondary emission from the anode can be reduced by special surface treatment.

An example for a large scale power plant are the tetrode amplifiers driving the main $200\unsty{MHz}$ travelling wave acceleration structures of the Super Proton Synchrotron (SPS) at CERN.
\begin{figure}[htb]
    \centering
	\includegraphics[height=0.349\linewidth]{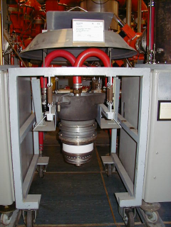}
	\includegraphics[height=0.349\linewidth]{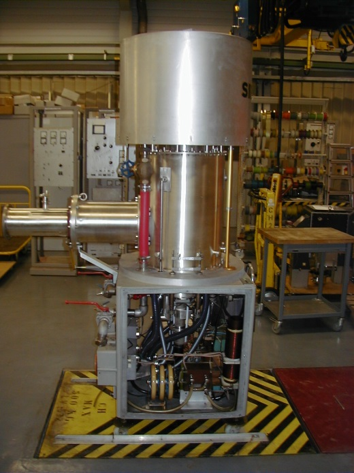}
	\includegraphics[height=0.349\linewidth]{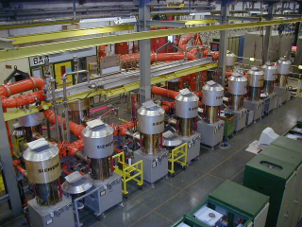}
	\caption{Tetrode amplifier driving part of the $200\unsty{MHz}$ RF system of the SPS at CERN~\cite{bib:montesinos2018}. Images courtesy of E. Montesinos.}
	\label{fig:TubeAmplifierExample}
\end{figure}
The water-cooled tube alone~(Fig.~\ref{fig:TubeAmplifierExample}, left) has an outer diameter of $255\unsty{mm}$ and weights about $35\unsty{kg}$~\cite{bib:siemens1994}. It is housed in an amplifier trolley which delivers an RF power of $125\unsty{kW}$~(Fig.~\ref{fig:TubeAmplifierExample}, center). The outputs of eight amplifier units are combined together to one power plant driving the accelerating structure through coaxial lines with a maximum peak power of $1\unsty{MW}$. Since 2020, the SPS is equipped with six accelerating structures, four of which are powered by tube-based amplifiers. The remaining two are driven by new solid state amplifiers (Fig.~\ref{fig:AmplifierSolidStateTowerExamples}, right). All amplifiers are located at the surface, some $60\unsty{m}$ above the accelerator tunnel.

\begin{figure}[htb]
    \centering
	\includegraphics[height=0.3\linewidth]{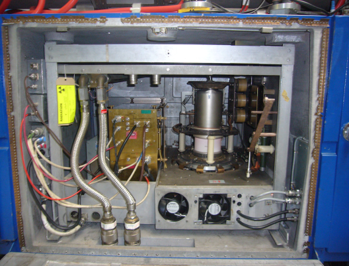} \hspace*{4em}
	\includegraphics[height=0.3\linewidth]{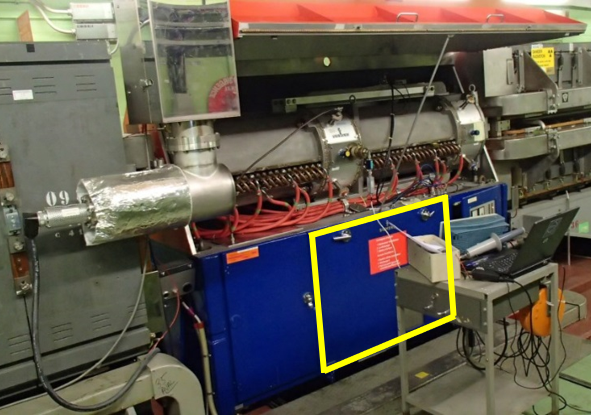}
	\caption{Amplifier trolleys for the PS $10\unsty{MHz}$ cavities~(left), and assembled directly below the RF cavity~(right).}
	\label{fig:TubeAmplifierExamplePS10MHz}
\end{figure}

The main accelerating cavities in the CERN Proton Synchrotron (PS) are driven by tetrode amplifiers as well~(Fig.~\ref{fig:TubeAmplifierExamplePS10MHz}).
To keep the delay between amplifier and cavity as short as possible, the amplifier unit is located directly below the RF cavity, as shown in Fig.~\ref{fig:TubeAmplifierExamplePS10MHz} (right, yellow indication). It can deliver about $60\unsty{kW}$ in the frequency range from $2.8\unsty{MHz}$ to $10\unsty{MHz}$. The tetrode amplifier is the obvious choice for this particular application. Due to the space constraints the high power density would be difficult to achieve with solid state devices. The tube itself is very compact and the voluminous power supplies are installed in auxiliary buildings. Additionally, the amplifier must be operated in a radioactive environment, just below the level of the beam pipe, leaving little spacing to shield a solid state amplifier.

\subsection{Linear beam tubes}

The mechanical dimensions of amplifier tubes scale with the power dissipation, the higher the output power the larger the device must be dimensioned to assure appropriate cooling. With increasing frequency the mechanical size of the tube therefore cannot be considered small compared to the RF wavelength any longer, and travelling wave effects must be taken into account. Beam tubes make use of these effects to achieve high gain and power output levels. The most prominent example of a linear beam tube is the klystron.

\begin{figure}[htb]
    \centering
	\includegraphics[height=0.5\linewidth]{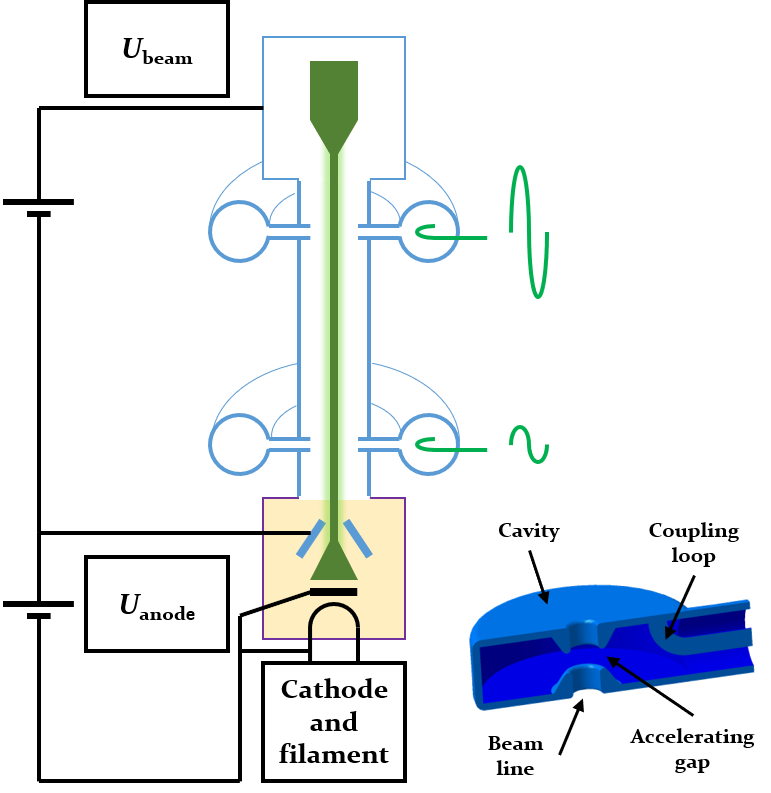}
	\caption{Sketch of a simple klystron amplifier, together with a typical coupling cavity~\cite{bib:montesinos2018}. Image courtesy of E.~Montesinos.}
	\label{fig:LinearBeamTubeKlystron}
\end{figure}

The simplified concept of the klystron, which can also be considered as a small electron accelerator in itself, is sketched in Fig.~\ref{fig:LinearBeamTubeKlystron}.
Again an electron cloud is generated by thermionic emission of electrons at a cathode. The electrons are accelerated by an anode in the so-called electron gun part of the klystron. They then pass through a drift region and are finally captured by a collector. At minimum two RF cavities are located in the drift space between electron gun and collector. The first one, near the electron gun, is driven by a low-power RF signal at the desired frequency. The resulting electric field in that cavity, the bunching cavity, accelerates or decelerates the electrons,  and it is hence generating a velocity modulation of the electron beam at the RF frequency. In the drift space between the two cavities, the velocity modulation leads to the formation of clusters of electrons. Faster electrons catch up to those in front of them, while slower electrons fall behind. At the arrival of the second cavity the electron beam is maximally bunched. The bunches are now exciting an RF signal in this output cavity, much larger than the input signal driving the klystron to generate the velocity modulation. Klystrons distinguish themselves by their high gain of the order of $60\unsty{dB}$.

\begin{figure}[htb]
    \centering
	\includegraphics[height=0.24\linewidth]{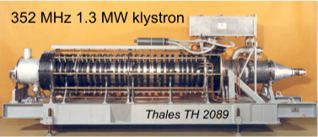}
	\includegraphics[height=0.24\linewidth]{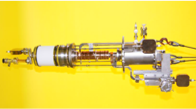}
	\caption{Klystron from the RF system of the Large Electron Positron (LEP collider) producing $1\unsty{MW}$ continuous output power at $352\unsty{MHz}$ (left). Body of a $12\unsty{GHz}$ pulsed klystron with $50\unsty{MW}$ peak power.}
	\label{fig:KlystronExampleLEPCLIC}
\end{figure}

Real klystrons are usually significantly more evolved than the simple example shown above. Additional cavities may be inserted in the drift space between in- and output resonators to assure optimum bunching. Solenoid magnets are moreover required around the drift region to keep the electron beam focused. About $50$ units of the large klystrons shown in Fig.~\ref{fig:KlystronExampleLEPCLIC} (left) were powering the RF system of the Large Electron Positron (LEP collider) to compensate the enormous synchrotron radiation losses.
The square box at the top right of the photo is the output waveguide. The pulsed klystrons at $12\unsty{GHz}$ for the Compact Linear Collider (CLIC) are much more compact. The cavities are the small copper structures visible in the central part of the klystron body. The peak power of such klystrons is about $50\unsty{MW}$, which is delivered during $1.5\unsty{\mu s}$ long pulses.

\subsection{Solid state amplifiers}

Compared to tubes, the output power per transistor device of a solid state amplifier is significantly lower. The active semiconductor die must be small, with the size limiting the maximum dissipated power. Internally connecting multiple dies in parallel within an RF transistor~\cite{bib:wood2008} already increases the output power, but also the internal capacities. Additionally, the thickness of the bonding wires, contacting the die to the external connection plates, is limited by the bond pad size which must be kept small to again minimize parasitic capacities~\cite{bib:dye1992}. These bonding wires hence cannot carry high currents. Solid state amplifiers for particle accelerators are therefore composed of a large number of amplifier modules, with an individual output power in the range of $1\unsty{kW}$ each, combined together to achieve the desired output power.

\begin{figure}[htb]
    \centering
	\includegraphics[height=0.4\linewidth]{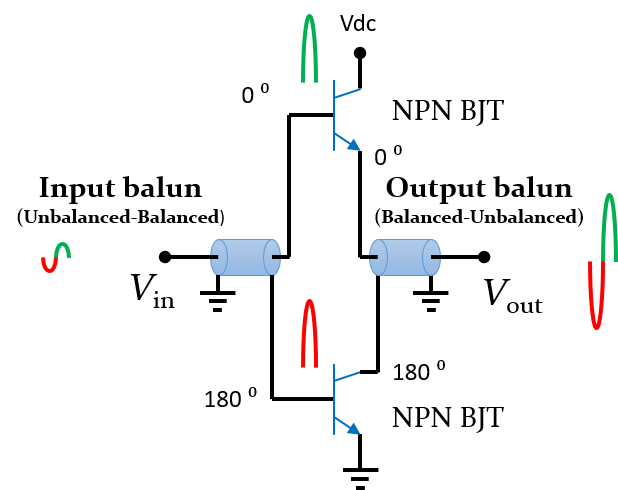}
	\caption{Basic configuration of a solid state RF power amplifier.}
	\label{fig:AmplifierSolidStatePrinciple}
\end{figure}

The conceptual schematics of an RF amplifier illustrating the basic building blocks is sketched in Fig.~\ref{fig:AmplifierSolidStatePrinciple}.
Auxiliary circuity as the working point control for the transistors is omitted for simplification. To achieve a reasonable compromise between efficiency and linearity the transistors are mostly operated in the so-called class AB. It mainly avoids excessive stand-by current by using each transistor as a switch with a fixed polarity. The sinusoidal input signal is split into positive and negative half wave, each of which are amplified separately and recombined at the output. In the audio frequency range complementary transistors with matched behaviour but opposite polarity are very common. However, such matched pairs with equal characteristics would be extremely difficult to produce at RF frequencies. In a limited frequency range it is technically much easier to swap the polarity of the RF signals and to amplify the signals by the same transistor type in both branches as illustrated in Fig.~\ref{fig:AmplifierSolidStatePrinciple}. These inversion transformers are called baluns, an acronym of balance-to-unbalanced and vice versa. The input signal (Fig.~\ref{fig:AmplifierSolidStatePrinciple}, red and green) is split by a balun transformer into positive (green) and inverted negative half waves. A second balun after the transistor stages recombines both half waves to the amplified original input signal.

\begin{figure}[htb]
    \centering
	\includegraphics[height=0.3\linewidth]{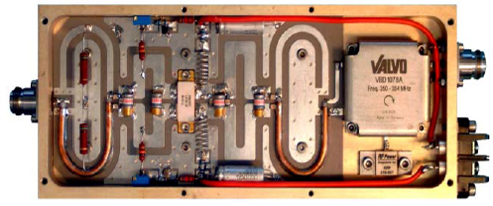}
	\caption{Example for a power amplifier modules according to the schematic diagram Fig.~\ref{fig:AmplifierSolidStatePrinciple}. This particular module, delivering an output power of $330\unsty{W}$ is part of the first generation solid state amplifiers of the synchrotron light source Soleil~\cite{bib:jacob2010, bib:jacob2016}.}
	\label{fig:AmplifierSolidStateExample}
\end{figure}

\begin{figure}[htb]
    \centering
	\includegraphics[height=0.352\linewidth]{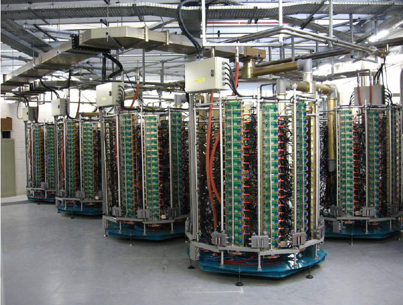}
	\includegraphics[height=0.352\linewidth]{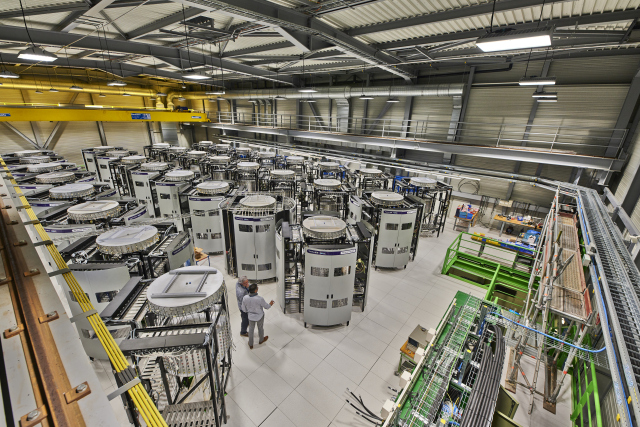}
	\caption{Solid state amplifier towers of Soleil (left) and SPS (right).}
	\label{fig:AmplifierSolidStateTowerExamples}
\end{figure}

Figure~\ref{fig:AmplifierSolidStateExample} shows an example amplifier module. 
The input signal is injected through the RF connector on the left. It passes through the input balun, a coaxial U-shaped line together with a transmission line on the printed circuit board and reaches the RF transistors, the rectangular ceramic body at the centre of the amplifiers. To achieve an optimum parameter match of both transistors, these are commonly even on the same semiconductor die and are therefore available as a single component. The output balun design is basically the same as at the amplifier input. To protect the individual module from reverse power, an optional circulator makes sure that any power injected through the output is safely dumped in a termination resistor, visible in Fig.~\ref{fig:AmplifierSolidStatePrinciple} just below the circulator box marked ``Valvo''. In total 64 of these modules are combined to an amplifier tower delivering an output power of $20\unsty{kW}$~(Fig.~\ref{fig:AmplifierSolidStateTowerExamples}), and several towers are re-combined to increase the output power even further.
The most powerful solid state amplifiers are presently the two $200\unsty{MHz}$ power plants for the SPS at CERN. In total $16$ amplifier towers, containing eighty $2\unsty{kW}$-modules each, are combined to deliver an output power of about $1.6\unsty{MW}$ per amplifier. \enlargethispage{\baselineskip} 

\subsection{Comparison}

\begin{figure}[htb]
    \centering
	\includegraphics[height=0.6\linewidth]{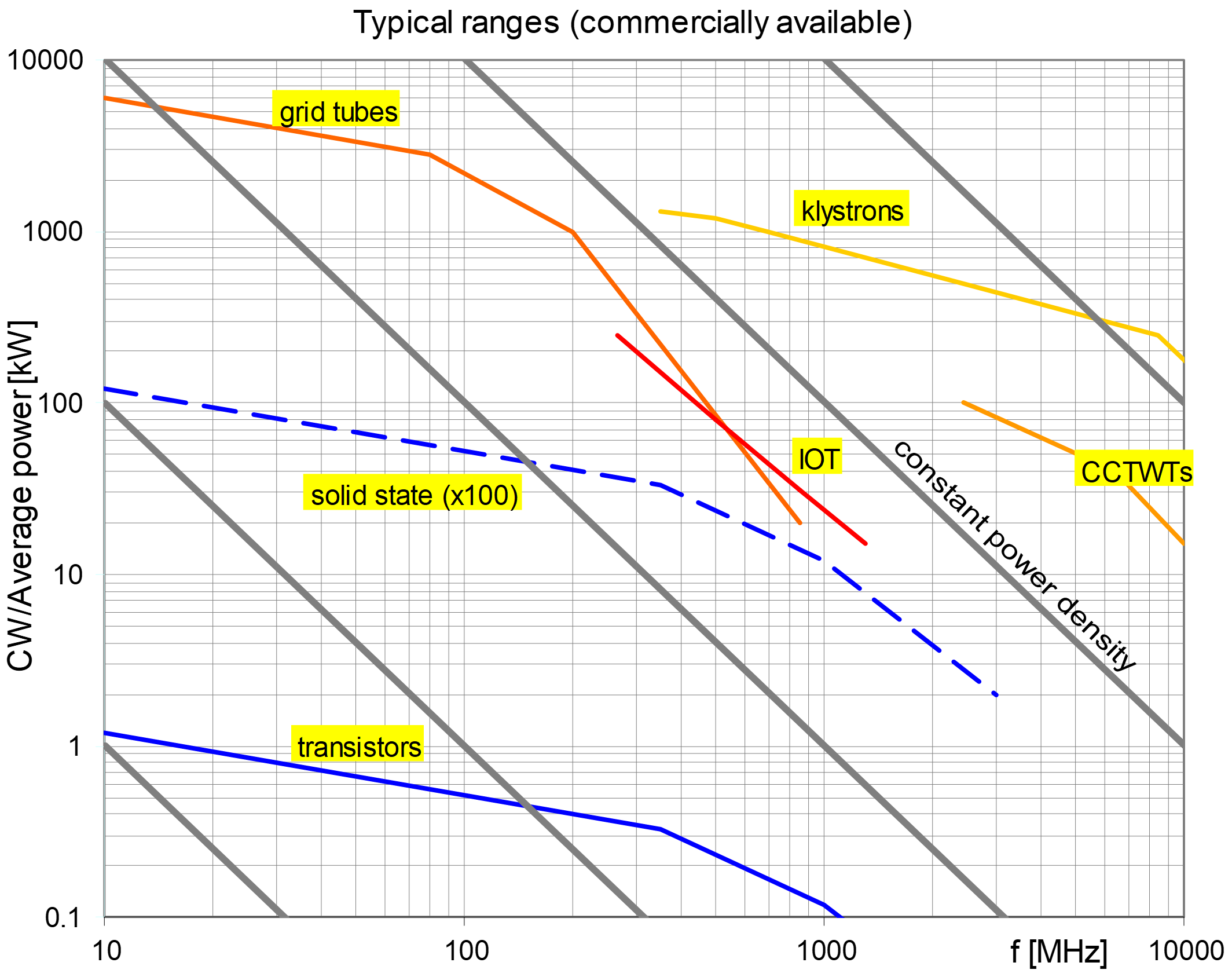}
	\caption{Comparison of commercially available amplifier types: Solid state transistor amplifiers, grid tubes (tetrodes and diacrodes), Inductive Output Tubes (IOT), klystrons and Coupled-Cavity Travelling Wave Tubes (CCTWT). Image courtesy of Erk Jensen (CERN), updated.}
	\label{fig:AmplifierPowerDevicesTableErk}
\end{figure}

The continuous wave power ranges of commercially available RF power sources are summarized in Fig.~\ref{fig:AmplifierPowerDevicesTableErk}.
The low and medium power regime is dominated by transistors. The combination of a large number of solid state amplifiers is required to match the large output power of tubes. While grid tubes, mainly the tetrodes, dominate the frequency range up to approximately $1\unsty{GHz}$, their physical size becomes to large compared to the wavelength at higher frequencies. Linear beam tubes like klystrons and coupled-cavity travelling wave tubes take over in the upper frequency range.

The choice of the amplifier technology for an RF system in a particle accelerator depends very much on the specific application. A non-exhaustive list of criteria and arguments is presented in Table.~\ref{tab:AmplifiertechnologyChoices}.
\begin{table}[htb]
	\caption{Guidelines for the choice of the RF amplifier technology.}
	\label{tab:AmplifiertechnologyChoices}
	\centering\small
	\begin{tabular}{ll}
		\hline\hline \\[-0.5em]
		\textbf{Prefer tube amplifiers, when} & \textbf{Prefer solid state amplifiers, when} \\[0.5em] \hline \\[-0.5em]
		
		\tabitem Amplifier must be installed & \tabitem Amplifier can be located \\
		\phantomtabitem the accelerator tunnel & \phantomtabitem in non-radioactive environment \\[0.5em]
		
		\tabitem Expecting important spikes from & \tabitem Optionally a circulator can be \\
		\phantomtabitem beam induced voltage & \phantomtabitem installed to protect the amplifier \\[0.5em]
		
		\tabitem Large output power of a single & \tabitem Delay due to unavoidable \\
		\phantomtabitem device is required, without combiners & \phantomtabitem combiner stages is no issue \\[0.5em]
		
		\tabitem Not much space is available & \tabitem Sufficient space can be made available \\[0.5em]

		\tabitem High peak power in pulsed mode & \tabitem Continuous operation preferred, \\
		& \phantomtabitem although pulsed amplifiers exist \\[0.5em]

		\tabitem Amplifier must be compact and/or & \tabitem Amplifier can be separate \\
		\phantomtabitem close to cavity & \phantomtabitem from the cavity \\[0.5em]
		
		\hline \hline

	\end{tabular}
\end{table}
The table only includes technical arguments relevant from the beam point of view, but intentionally excludes those related to maintainability, reliability, lifetime or cost.

The cost to provide and operate the technical infrastructure in terms of electrical power supply, cooling, etc. can become very important for high-power amplifiers and should nonetheless be taken into for the choice. For example the high-voltage installations for tube-based power plants require special care and intensive maintenance, while solid state amplifiers work with moderate supply voltages which are easier to handle.

Additionally, the overall efficiency of an RF power plant can contribute significantly to the cost of operation. For a fair comparison it must therefore include the total power lost in the entire infrastructure needed to run an installation.

For many applications, for example the power amplifier in conventional synchrotron light sources, both amplifier technologies can be considered as viable options. In other case, like the installation of an RF amplifier in a radioactive environment, the tube technology is better suited.

\section{Low-level RF system and feedback}

The amplifier powering an accelerating cavity is itself driven, at a low power level, by the low-level RF system~\cite{bib:garoby1991}. It makes sure that the correct frequency at the right phase at the intended amplitude is generated. In synchrotrons these signals are actually derived by regulation loops, referred to as feedback systems and longitudinal beam control, to optimally accelerate the circulating beam.

As mentioned above, most high-energy particle accelerators are equipped with multiple RF stations to manipulate the beam. A distinction is therefore made between local and global feedback loops. The local loops act on each individual RF station separately, without any knowledge of signals from other RF stations. Global feedback loops are those which treat all RF stations just as one single large RF system, and they hence act globally on all RF stations simultaneously.

\subsection{Local feedback systems}

The energy transfer between an RF cavity and the beam occurs in both directions. Both, the RF amplifier as well as the beam can induce voltage in the cavity. Of course, this beam induced voltage, causing energy transfer from the cavity back to the beam, is unintended and must be reduced.

\begin{figure}[htb]
    \centering
	\includegraphics[height=0.2\linewidth]{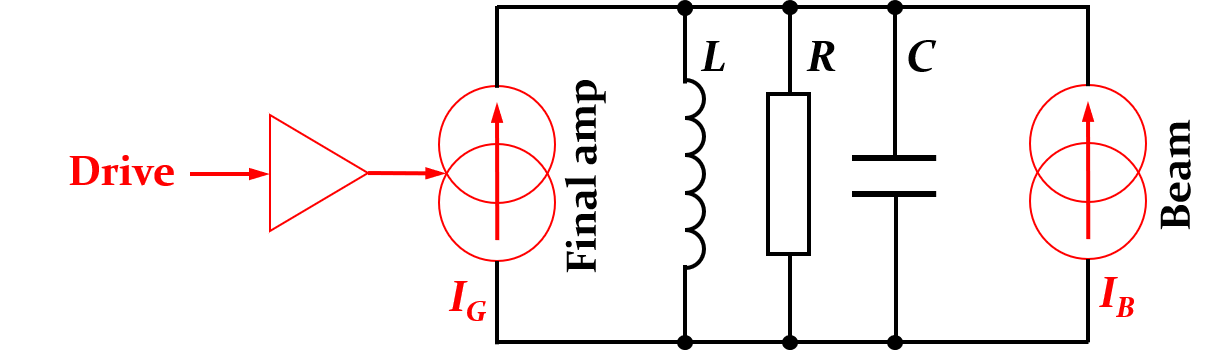}
	\caption{Lumped element circuit model of a cavity driven simultaneously driven by amplifier and beam.}
	\label{fig:LumpedElementModelBeamLoadingNoFeedback}
\end{figure}

A simplified lumped element circuit is sketched in Fig.~\ref{fig:LumpedElementModelBeamLoadingNoFeedback}.
In this model amplifier and beam can be represented as current sources. To reduce the beam-induced voltage across the cavity gap one could simply reduce the shunt impedance, $R$ of the resonator. However, this would be very inefficient for obvious reasons, as the power amplifier has to provide a higher power as well, which translates not only in wasted power but also in a significant cost increase for the amplifier and its operation. According to $P = V^2/(2R)$, the power is directly proportional to the reduction of the beam-induced voltage, $V_\mathrm{ind} = R \cdot I_\mathrm{B}$ due to a decreased shunt impedance. \enlargethispage{\baselineskip} 

\begin{figure}[htb]
    \centering
	\includegraphics[height=0.2\linewidth]{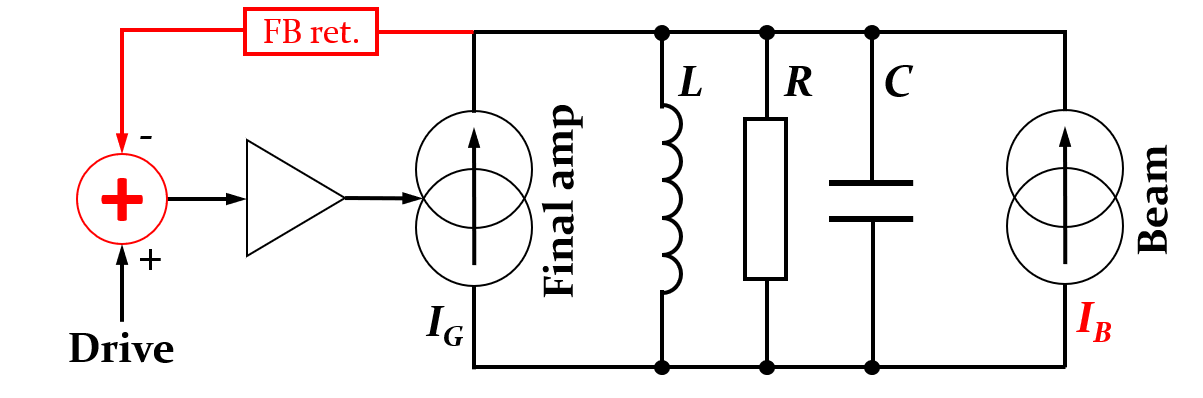}
	\caption{Lumped element circuit model of an RF system with direct feedback.}
	\label{fig:LumpedElementModelBeamLoadingFeedback}
\end{figure}

A more more elegant way to decrease the apparent impedance for the beam can be realized by using the power amplifier to counteract the beam-induced voltage. The simplified lumped element circuit model is shown in Fig.~\ref{fig:LumpedElementModelBeamLoadingFeedback} for such a configuration with direct feedback.
The drive signal to the cavity, which does not include any beam induced contribution, is compared to a signal picked-up at the cavity gap. The latter contains the sum voltage driven by amplifier and beam. The difference between both signals, the beam-induced signal, is inverted and amplified. The equivalent impedance, $Z_\mathrm{eq}$ with the direct feedback system can be calculated by evaluating the differential gap voltage change, $dV$ for a differential beam current change, $dI_\mathrm{B}$ and becomes~\cite{bib:baudrenghien2005a}
\begin{equation*}
    Z_\mathrm{eq} = \frac{dV}{dI_\mathrm{B}} = \frac{ Z(\omega) }{ 1 + g_\mathrm{OL} } \, ,
\end{equation*}
where $Z(\omega)$ represent the impedance of the bare resonator according to~Eq.~(\ref{eqn:lumpedElementResonanceImpedance}). The open loop gain, $g_\mathrm{OL}$ describes the transfer function of the entire feedback loop, from the cavity gap through the return path and amplifier back to the cavity gap. The full shunt impedance of the cavity hence remains visible to the amplifier, while it is reduced by $1/( 1 + g_\mathrm{OL})$ for the beam trying to induce voltage into the cavity. 

\begin{figure}[htb]
    \centering
	\includegraphics[height=0.3\linewidth]{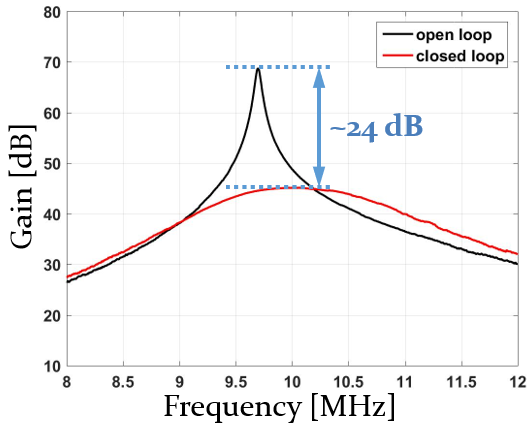}
	\caption{Measured transfer function of a cavity with and without direct feedback. Measurement courtesy of G. Favia.}
	\label{fig:LumpedElementModelBeamLoadingFeedbackExample}
\end{figure}

Measured transfer functions are plotted in Fig.~\ref{fig:LumpedElementModelBeamLoadingFeedbackExample} for the open (black) and closed (red) loop case for the example of the main accelerating cavities in the CERN PS.
The effective impedance reduction of the cavity in this particular case is about $24\unsty{dB}$, corresponding to an about $16$ times smaller beam induced voltage than without the direct feedback system. Although the amplifier is installed directly below the RF cavity to keep the delay as well as the associated phase rotation as small as possible, it is this delay which, in combination with the large bandwidth, limits the maximum feedback gain.

As the de-phasing $\Delta \phi = \Delta \omega \tau$ is proportional to the frequency offset with respect to the resonance frequency, intentionally reducing the feedback bandwidth is an option to achieve larger feedback gain. One example is introduced in the next section.

\subsection{1-turn delay feedback}

The longitudinal beam spectrum in a synchrotron is concentrated around multiples of the revolution frequency, $n f_\mathrm{rev}$. Any longitudinal impedance is hence only excited at these frequencies, which can be exploited for the 1-turn delay feedback concept. This feedback system implements a periodic comb filter with high gain only in very narrow frequency bands around the revolution frequency harmonics. Thanks to this significantly smaller bandwidth of each pass-band compared to the overall cavity transfer function~(Fig.~\ref{fig:LumpedElementModelBeamLoadingFeedbackExample}), additional impedance reduction at the revolution frequency harmonics is obtained well beyond the stability limit of a wideband feedback system. The unavoidable latency introduced by the signal processing is compensated by applying the correction with a delay of one turn.

\begin{figure}[htb]
    \centering
	\includegraphics[height=0.3\linewidth]{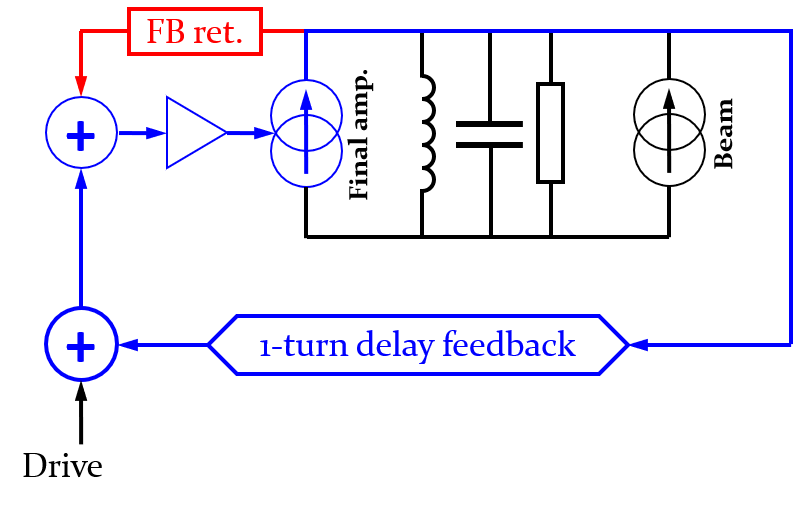}
	\caption{Lumped element circuit model of an RF system with direct and 1-turn delay feedback.}
	\label{fig:LumpedElementModelBeamLoading1TturnDelayFeedback}
\end{figure}

\begin{figure}[htb]
    \centering
	\includegraphics[height=0.255\linewidth]{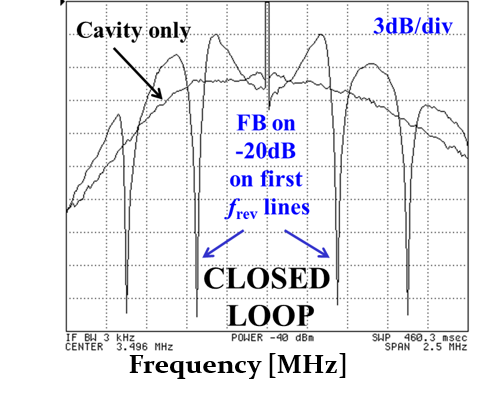} \hspace*{2em}
	\includegraphics[height=0.26\linewidth]{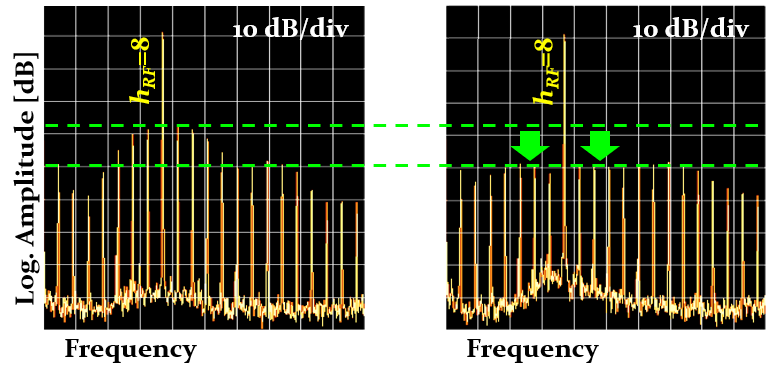}
	\caption{Comparison of a cavity transfer function without and with 1-turn delay feedback. The two measurements of the beam induced spectrum at the cavity gap (middle, right) demonstrate that the relevant revolution frequency harmonics in the bandwidth of the cavity resonance are well reduced by 1-turn delay feedback. For the transfer function measurement without beam~(left), the gain was pushed to~$20\unsty{dB}$, while the beam measurements were recorded with more conservative settings (about $12\unsty{dB}$ gain). Images courtesy of D.~Perrelet.}
	\label{fig:1TurnDelayFeedbackOpenClosedLoop}
\end{figure}

The overall set-up of a cavity with direct and 1-turn delay feedback is sketched in Fig.~\ref{fig:LumpedElementModelBeamLoading1TturnDelayFeedback}. 
Due to the total loop delay of the revolution period, $T_\mathrm{rev}$, the hardware can be located far away from the cavity, outside the accelerator tunnel. The resulting modification of the transfer function is illustrated in Fig.~\ref{fig:1TurnDelayFeedbackOpenClosedLoop}~(left).
With the additional feedback loop active, deep narrow-band notches counteract any beam-induced voltage at the revolution frequency harmonics. In between these frequencies the impedance is actually even larger with than without the feedback. At these frequencies however, the longitudinal beam spectrum has no content and even the enlarged impedance cannot be excited by the beam. The beneficial effect of the 1-turn delay feedback is also demonstrated by measuring the spectrum of the beam induced voltage at the cavity gap~(Fig.~\ref{fig:1TurnDelayFeedbackOpenClosedLoop}, middle and right). With feedback (right) the revolution frequency harmonics within the cavity bandwidth are reduced well beyond the reach of the direct, wide-band feedback loop. This two-staged approach of direct and 1-turn delay feedback systems to reduce beam induced voltage makes clever usage of the beam periodicity in a synchrotron. . \enlargethispage{\baselineskip} 

\subsection{Global feedback systems}

\begin{figure}[htb]
    \centering
	\includegraphics[height=0.33\linewidth]{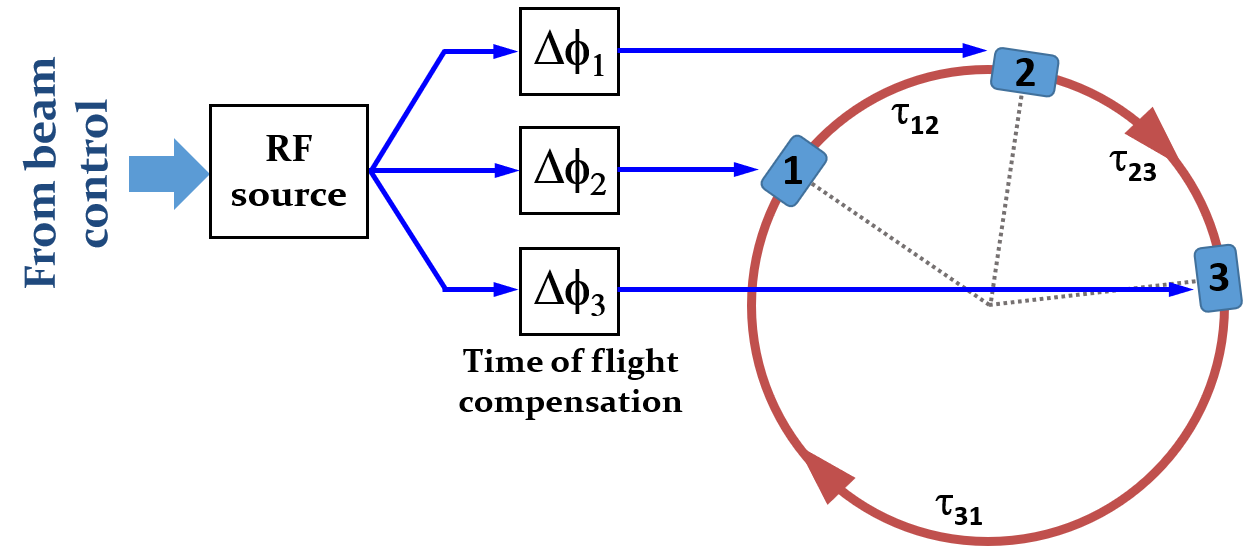}
	\caption{Model of a synchrotron with three RF stations. The relative phases must be adjusted such that the time of flight of the beam is compensated such that they are all in phase at the moment of the beam passage.}
	\label{fig:GlobalFeedbackDriveCavitiesInUnison}
\end{figure}

Global feedback systems act on all RF stations simultaneously and hence drives the entire set of them just like a single RF station. The global loops are also referred to as the longitudinal beam control~\cite{bib:boussard1973,bib:baudrenghien2005b}. A simplified model of an accelerator with three RF stations is sketched in Fig.~\ref{fig:GlobalFeedbackDriveCavitiesInUnison}. 
To compensate the time of flight of the beam from RF station to RF station, a de-phasing is applied. It assures that all stations are in phase at the moment of the beam passage. The frequency of the RF source is sweeping with the increasing beam velocity, but the relative phases are static as the geometrical azimuth between the RF stations is constant. Global feedback loops apply small corrections in frequency and phase to the RF signal generated by the common master RF source. 

Distinction is made between three main global feedback loops for synchrotrons. Firstly, the beam phase loop locks the phase of the RF system to the centre the phase of the circulating bunches. It damps any common-mode dipole oscillations. However, the phase loop introduces a slow drift in average RF frequency which leads to a radial displacement if uncorrected. This is avoided by either a radial loop or a synchronization loop. The radial loop measures the radial displacement of the beam and corrects it by changing the RF frequency, $f_\mathrm{RF} = h \beta c/(2 \pi R)$. The synchronisation loop locks the RF frequency to an external reference frequency. Additionally, at the beam transfer from one synchrotron to the next, it is necessary to lock the RF frequencies of the two rings. This scenario is also handled by a synchronization loop.

\subsubsection{Beam phase loop}
\label{sec:BeamPhaseLoop}

The beam phase loop locks the phase of the accelerating wave with respect to the circulating bunches such that the relative phase between a longitudinal beam signal and the RF voltage in the cavities is set to the synchronous phase. In this case the beam cannot escape the accelerating wave and any bunch phase oscillations, with all bunches oscillating in phase, are efficiently suppressed.

\begin{figure}[htb]
    \centering
	\includegraphics[height=0.21\linewidth]{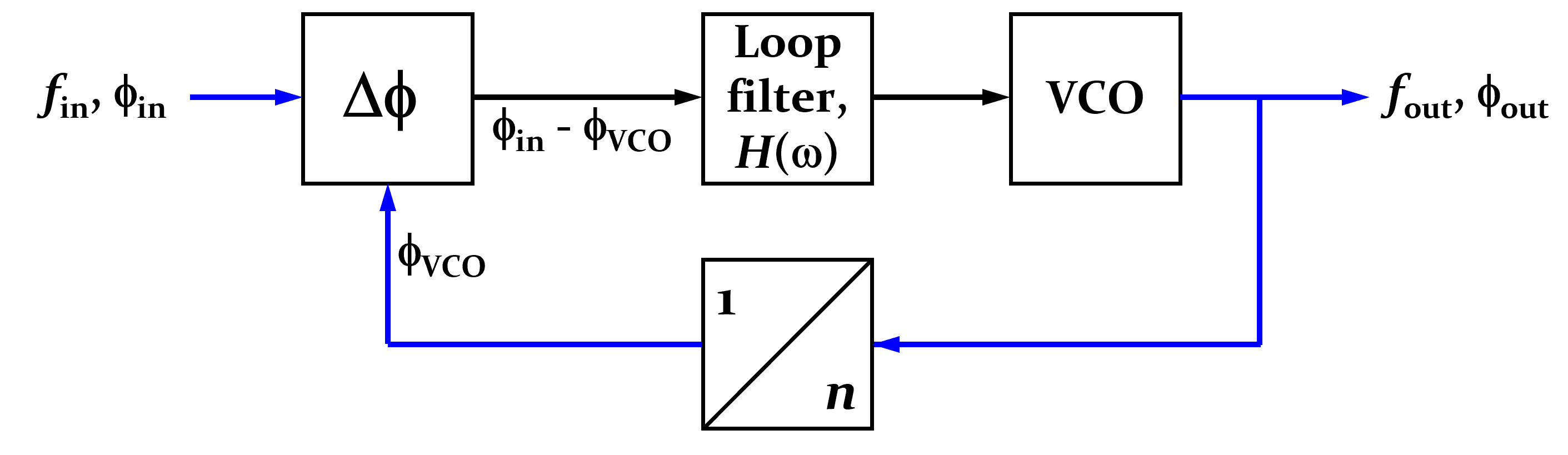}
	\caption{Block diagram of an electronic phase-locked loop (PLL). RF signals are represented in blue, while connections marked in black are non-RF signals at low frequency.}
	\label{fig:PhaseLockedLoopBlockDiagram}
\end{figure}

The beam phase loop is a particular variant of the electronic phase-locked loop (PLL), which should be briefly introduced here. In telecommunication technology, the PLL is mainly employed for frequency regeneration and multiplication, as well as for frequency synthesis. Figure~\ref{fig:PhaseLockedLoopBlockDiagram} shows a block diagram of a simple PLL.
It locks phase and frequency at the output, $f_\mathrm{out}$ and $\phi_\mathrm{out}$, to the corresponding signal at the input at $f_\mathrm{in}$ and $\phi_\mathrm{in}$. The voltage controlled oscillator (VCO) generates an output frequency,~$\omega_\mathrm{VCO}$ proportional to its input voltage, $V_\mathrm{in}$ according to
\begin{equation*}
    \omega_\mathrm{VCO} = 2 \pi f_\mathrm{VCO} = \frac{d\phi}{dt} = K_\mathrm{VCO} V_\mathrm{in} \, .
\end{equation*}
The voltage-to-frequency scaling is given by $K_\mathrm{VCO}$. The division ratio, $n$ of the divider becomes the frequency multiplication factor. The output of the VCO, which is also the output of the PLL, passes through a frequency divider and is then compared in phase with the input signal. The phase difference is filtered and fed back to the input of the VCO. This closes the loop which, in locked state, assures that
\begin{equation*}
    \phi_\mathrm{out}/n - \phi_\mathrm{in} = \mathrm{const.} \hspace{1em} \mathrm{and} \hspace{1em} f_\mathrm{out} = n \cdot f_\mathrm{in} \, .
\end{equation*}
In this configuration the PLL acts as a frequency multiplier with a well-defined fixed phase relationship between in- and output signals.

\begin{figure}[htb]
    \centering
	\includegraphics[height=0.5\linewidth]{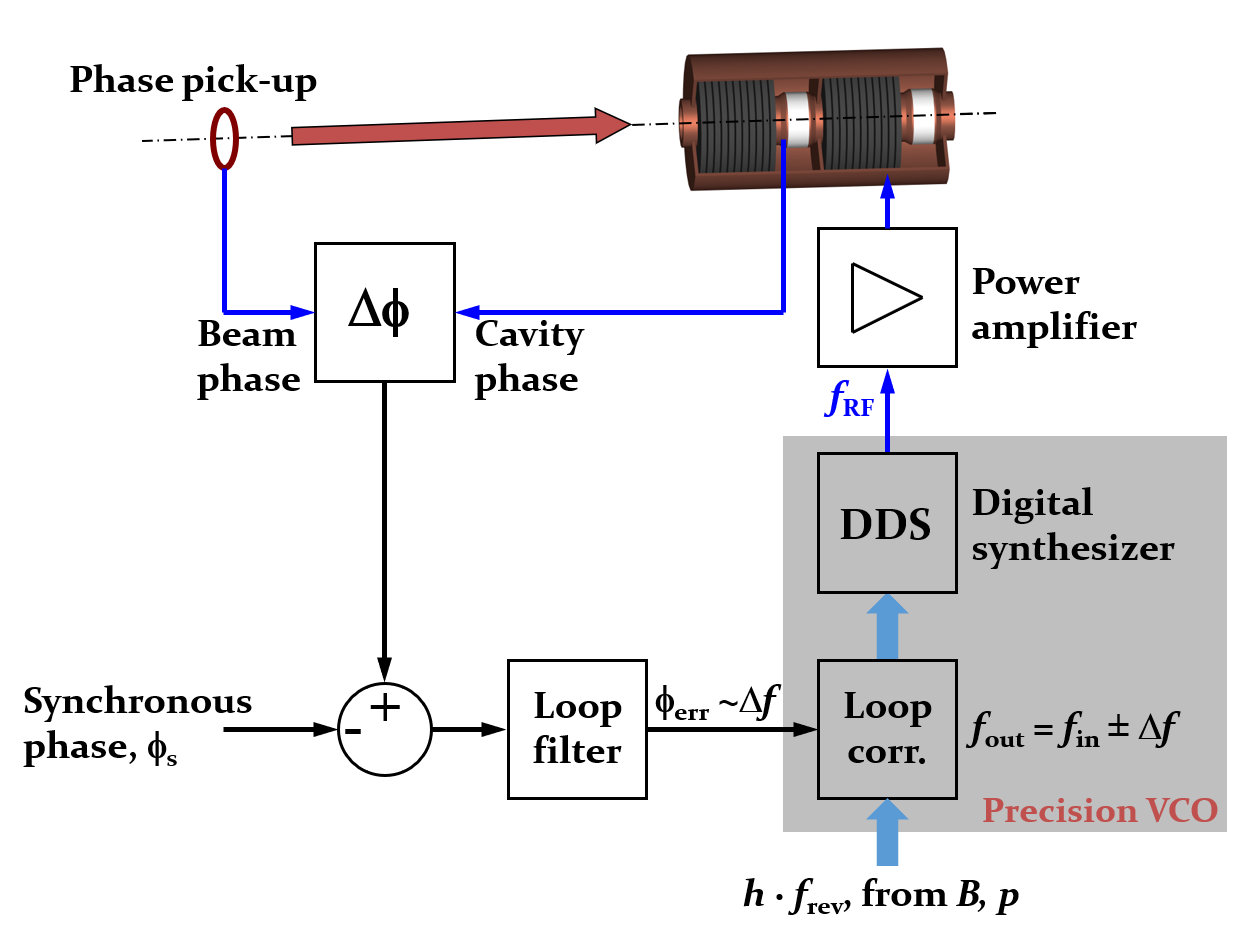}
	\caption{Block diagram of a beam phase loop in a hadron synchrotron. RF signals are represented in blue, while connections marked in black are non-RF signals at low frequency. Signals typically implemented digitally are shown as thick light blue arrows.}
	\label{fig:BeamPhaseLoop}
\end{figure}

Replacing the input signal by a signal from the circulating beam, and fixing the division ratio to unity, transforms the electronic PLL into a beam phase loop~(Fig.~\ref{fig:BeamPhaseLoop}).
The phase of the RF voltage in the accelerating cavity is compared to the reference phase from a longitudinal beam pick-up. After correction of the stable phase, $\phi_\mathrm{S}$ the phase difference is passed through a loop filter and then controls the frequency of a precision VCO to generate the drive signals for the power amplifiers. Only one single RF station is shown for simplicity, but more could be added following the scheme presented in Fig.~\ref{fig:GlobalFeedbackDriveCavitiesInUnison}.

To improve the performance of the loop and to cope with the sweeping revolution frequency acceleration, the VCO is implemented in the form of a precision RF source. An estimate of the expected revolution and RF frequency is numerically calculated from the bending field or momentum. This open loop frequency value is corrected by a small amount with the output from the loop filter~(Fig.~\ref{fig:BeamPhaseLoop}). The corrected RF frequency, then called closed-loop RF frequency, drives a direct digital synthesizer (DDS) RF source. The latter is typically a numerically controlled oscillator with perfect linearity and generates the RF signal at the corrected, closed-loop RF frequency used to drive the power amplifier. This closes the loop such that the phase and frequency of the RF system are locked to the phase of the circulating beam.

In modern beam control systems the entire loop may be implemented digitally, with analog-to-digital converters acquiring beam and cavity return signals. Conceptually identical to the set-up in Fig.~\ref{fig:BeamPhaseLoop}, digital loops feature better reproducibility, long-term stability and more flexibility for the dynamic configuration of the loop filter, also during the acceleration cycle.

\subsubsection{Radial loop}

The beam phase loop tracks the beam in terms of phase. However, it may induce unwanted drifts of the average frequency during acceleration. Again the beam itself can be taken as a reference to control the average RF frequency by monitoring its radial displacement with respect to the central orbit.

\begin{figure}[htb]
    \centering
	\includegraphics[height=0.25\linewidth]{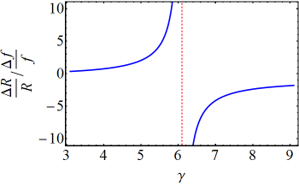}
	\caption{Dependence of radial displace with versus frequency near transition energy. In this example the transition occurs at $\gamma_\mathrm{tr} = 6.1$.}
	\label{fig:BeamRadialLoopPositionSensitivityNearTransition}
\end{figure}

The radial loop becomes essential around the transition energy. This is illustrated using the differential relation between frequency and radial offset~\cite{bib:bovet1970} 
\begin{equation*}
    \frac{\Delta f}{f} = \frac{\gamma_\mathrm{tr}^2-\gamma^2}{\gamma^2} \frac{\Delta R}{R} + \frac{1}{\gamma^2} \frac{dB}{B} \, ,
\end{equation*}
where $\gamma_\mathrm{tr}$ is the energy factor, $\gamma = E/E_0$. For a magnetic cycle with a slowly increasing bending field. the second term can be neglected. At transition energy, $\gamma = \gamma_\mathrm{tr}$, the effect of a longer or shorter path due to a higher or lower particle energy on the revolution frequency is compensated by the higher or lower velocity of the particle, and the revolution frequency is independent of the particle energy. On the contrary, the sensitivity of a relative radial displacement, $\Delta R/R$, with respect to a normalized frequency error, $\Delta f /f$ according to
\begin{equation}
    \frac{\Delta R}{R} = \frac{\gamma^2}{\gamma_\mathrm{tr}^2 - \gamma^2} \frac{\Delta f}{f} \, ,
    \label{eqn:radialOffsetVersusFrequencyOffset}
\end{equation}
shows that near transition energy even a tiny frequency error can cause a large radial displacement, sufficient to drive the beam horizontally into the vacuum chamber, see Fig.~\ref{fig:BeamRadialLoopPositionSensitivityNearTransition}.

\begin{figure}[htb]
    \centering
	\includegraphics[height=0.5\linewidth]{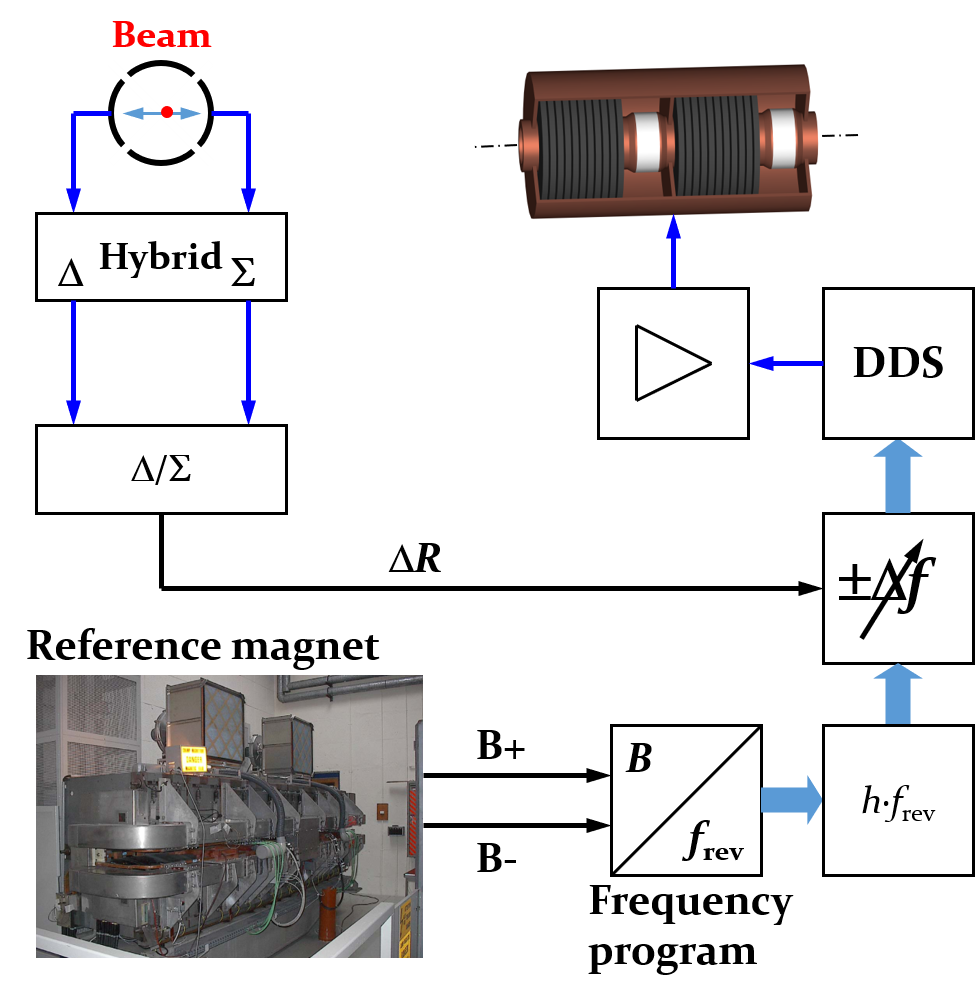}
	\caption{Block diagram of a radial loop to control the average RF frequency. The measurement of the bending field is performed in a reference magnet, located outside the accelerator tunnel but connected in series with the chain of bending magnets. The frequency program then converts this measured field into the open loop revolution frequency. RF signals are represented in blue, while connections marked in black are non-RF signals at low frequency. Signals typically implemented digitally are shown as thick light blue arrows.}
	\label{fig:BeamRadialLoop}
\end{figure}

The radial loop prevents this to happen by monitoring the radial displacement of the beam. A simplified configuration of such a control loop is sketched in Fig.~\ref{fig:BeamRadialLoop}.
The signals from the plates of a horizontal beam pick-up are converted to sum, $\Sigma$ and difference, $\Delta$ signals. The ratio between the two, $\Delta/\Sigma$ results in a measure proportional to the radial displacement of the beam. This displacement is translated into a correction of the frequency program, $f(B)$ derived from the bending field or a momentum program. The average RF frequency is hence controlled such that, according to Eq.~(\ref{eqn:radialOffsetVersusFrequencyOffset}), the radial displacement is minimized during acceleration and in particular around transition crossing.

\subsubsection{Synchronization loop}

Although essential to accelerate a beam through the transition energy, the radial loop has the major drawback of any other beam-based regulation loop. The measurement of the beam signal induces unwanted noise into the RF frequency within the loop and generates additional phase jitter of the RF signal at the cavity gap. Such a phase jitter degrades the longitudinal beam quality as it results in uncontrolled blow-up of the longitudinal emittance.

When the transition energy is not located between injection and extraction energy, and is therefore not crossed during the acceleration, the radial loop is better replaced by a synchronization loop to control the average revolution and RF frequencies. The combined block diagram of beam phase and synchronization loops is summarized in Fig.~\ref{fig:BeamPhaseAndSynchroLoop}.
\afterpage{
\vspace*{\fill}
\begin{figure}[!htb]
    \centering
	\includegraphics[height=0.5\linewidth]{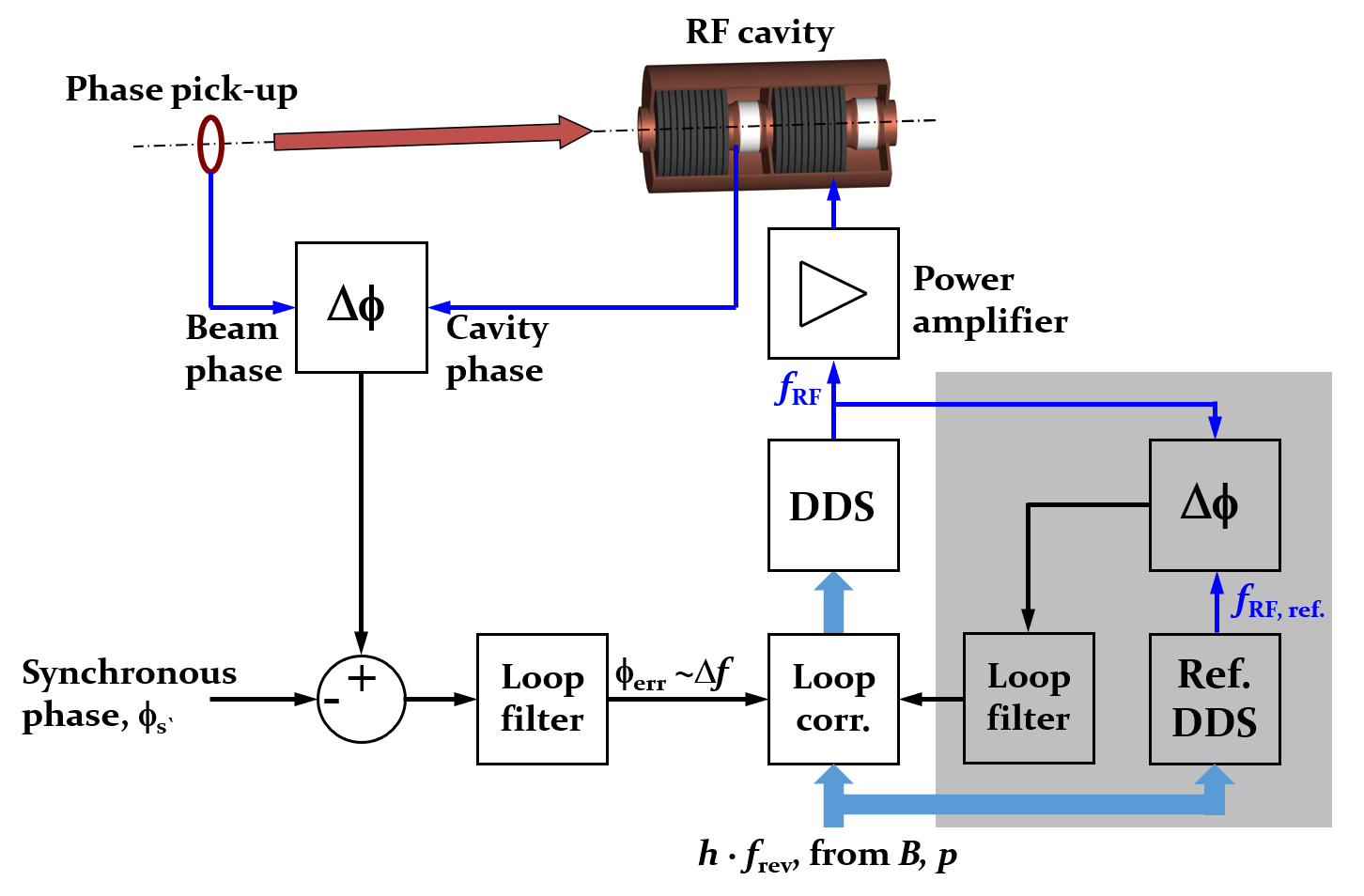}
	\caption{Block diagram of beam phase and synchronization loops. The grey insert shows the additional components of the synchronization loop to control the average RF frequency. RF signals are represented in blue, while connections marked in black are non-RF signals at low frequency. Signals typically implemented digitally are shown as thick light blue arrows.}
	\label{fig:BeamPhaseAndSynchroLoop}
\end{figure}
\begin{figure}[!htb]
    \centering
	\includegraphics[height=0.5\linewidth]{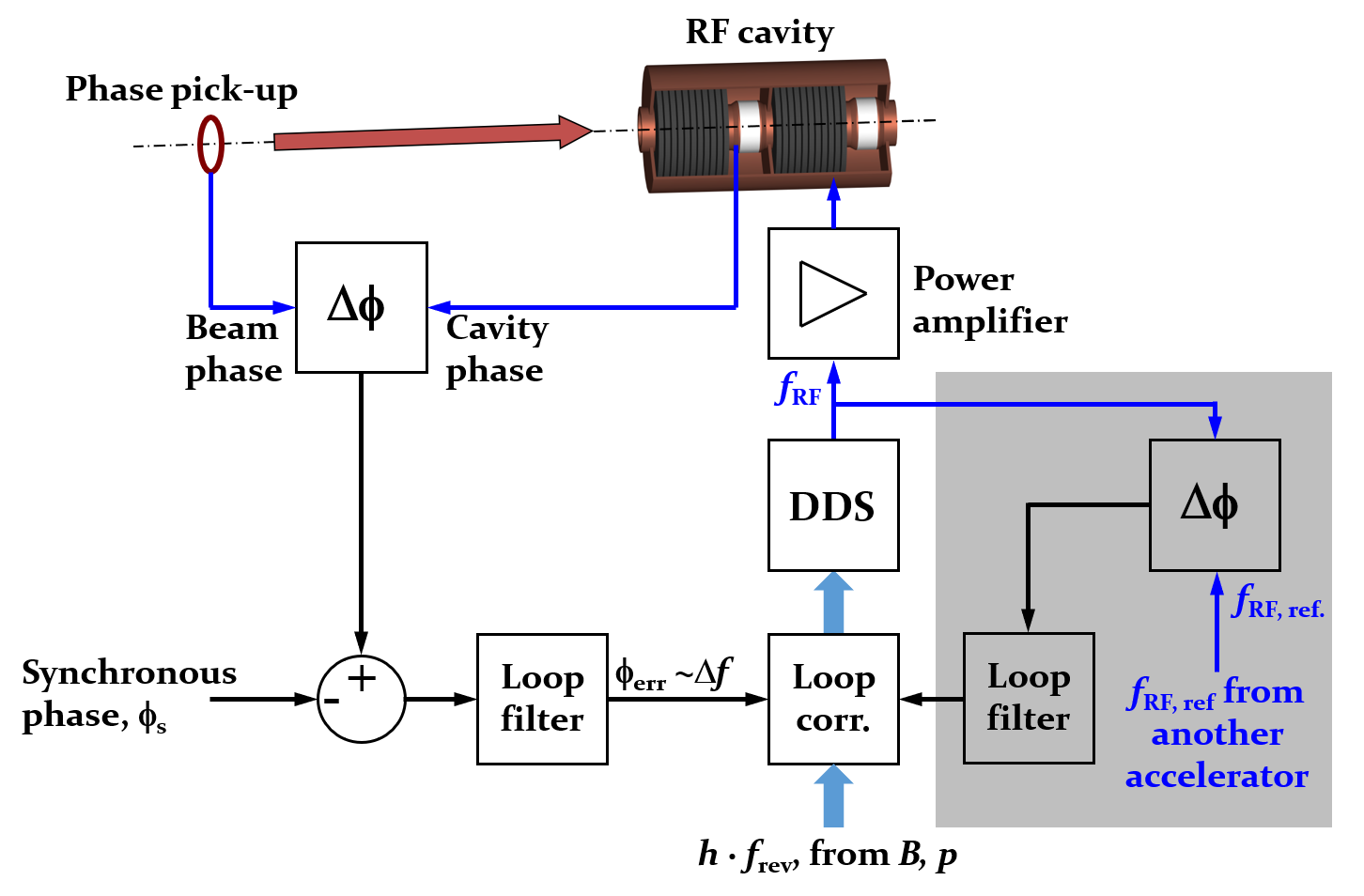}
	\caption{Block diagram of beam phase and synchronization (grey insert) loops, controlled by an external reference. RF signals are represented in blue, while connections marked in black are non-RF signals at low frequency. Signals typically implemented digitally are shown as thick light blue arrows.}
	\label{fig:BeamPhaseAndSynchroLoopExternalReference}
\end{figure}
\vspace*{\fill}
\clearpage
}
The beam phase loop has already been introduced in Sec.~\ref{sec:BeamPhaseLoop}. An additional reference synthesizer (Fig.~\ref{fig:BeamPhaseAndSynchroLoop}, reference DDS) generates an RF signal at the intended average RF frequency, which may be derived from the bending field or from a programmed frequency function. This reference frequency is compared in phase with the actual closed-loop RF frequency after the correction by the beam phase loop. The phase difference between the actual RF frequency, $f_\mathrm{RF}$ and the reference frequency, $f_\mathrm{RF,reference}$ passes a filter (grey inset), typically a low-pass with a cut-off frequency well below the synchrotron frequency to avoid any interaction with the much faster phase loop. However, the closed synchronization loop assures that the average RF frequency for the acceleration of the beam corresponds to the reference frequency, while at the same time the beam phase loop still tracks the phase of the circulating bunches.

The same loop can also be used to synchronize two synchrotrons with respect to each other. For a bunch-to-bucket transfer between two synchrotrons, the RF frequencies should be ideally identical in both or, more generally, integer harmonics of each other. In this case the internal reference synthesizer is replaced by an external RF frequency provided by the receiving synchrotron (Fig.~\ref{fig:BeamPhaseAndSynchroLoopExternalReference}) to which the average RF frequency and phase will be locked. The synchronization chain can be described as follows. 
\enlargethispage{-\baselineskip}
\begin{enumerate}
    \item In the receiving synchrotron the circulating bunches are locked to the RF voltage by its beam phase loop.
    \item The RF signals, together with a divided revolution frequency marker are transmitted as reference to the sending synchrotron.
    \item The average frequency and the phase in the sending synchrotron are locked to this external reference prior to the transfer.
    \item The beam phase loop of the sending synchrotron assures the phase lock between beam and RF voltage.
\end{enumerate}
In summary, thanks to the synchronization loop and the beam phase loops, the circulating bunches in both sending and receiving accelerator are locked in phase and frequency with respect to each other. In case of an integer circumference and revolution frequency ratio, $n$ of two accelerators, the beam in the smaller one will just make $n$ turns during one turn of the larger one. The overall system nonetheless has a fundamental periodicity with the revolution frequency of the larger accelerator.

More evolved schemes are required to transfer beams between two synchrotrons with a rational circumference ratio, but in most cases the synchronization loop plays a key role for the bunch-to-bucket transfer.

\section{Conclusion}

RF systems are integral parts of every linear and circular particle accelerator. Their parameters must be well chosen at the initial design stage, based on the requirements and constraints defined by the beam. This does not only assure the acceleration process, but also allows to manipulate the longitudinal beam parameters like, for example, bunch length, peak current or number of bunches. RF system are composed of the cavity which directly interact with the particle beam, and a set of low-level, as well as high-power electronics to provide sufficient RF voltage at the correct frequency and phase to the beam. As a major difference compared to most other components installed in accelerators, the interaction between beam and RF system is bidirectional and the beam, through induced voltage, influences the behaviour of the RF as well as vice versa. For the beam control loops, the beam actually becomes an integral part of the RF system and defines itself the optimum parameters for its acceleration.

\section{Acknowledgement}

The author would like to thank Maria-Elena Angoletta, Philippe Baudrenghien, Thomas Bohl, Giorgia Favia, J\"orn Jacob, Erk Jensen, John Molendijk, Gerry McMonagle, Mauro Paoluzzi, Damien Perrelet, Bernhard Schriefer, Lukas Stingelin, Fumihiko Tamura, Frank Tecker, Daniel Valuch for discussions and material provided in preparation of this course. Fruitful discussions with Thomas Bohl and Eric Montesinos, who also provided an important number of corrections and improvements to the manuscript, are particularly acknowledged.\enlargethispage{\baselineskip}


\begin{thebibliography}{99}

    \bibitem{bib:boussard1989} D.~Boussard, J.~M.~Brennan, T.~P.~R.~Linnecar, \emph{Fixed Frequency Acceleration in the SPS}, CERN SPS/89-49(ARF), CERN, Geneva, Switzerland, 1989, available at the URL \url{http://cds.cern.ch/record/204684}.
    
    \bibitem{bib:pirkl2005} W.~Pirkl, \emph{Choice of RF frequency}. In: CERN Accelerator School: Radio Frequency Engineering, Seeheim, Germany, 2005, available at the URL \url{http://cds.cern.ch/record/865925}.
    
    \bibitem{bib:kilpatrick1957} W.~D.~Kilpatrick, \emph{Criterion for Vacuum Sparking Designed to Include Both rf and dc}, Review of Scientific Instruments \textbf{28}, 824 (1957), avilable at the URL \url{https://doi.org/10.1063/1.1715731}.
    
    \bibitem{bib:wang1997} J~.W.~Wang and G.~A.~Loew, \emph{Field Emission and RF Breakdown in High-Gradient Room-Temperature Linac Structures}, SLAC-PUB-7684, SLAC, Stanford, Californa, USA, 1997, available at the URL \url{https://inspirehep.net/literature/454313}.
    
	\bibitem{bib:sands1970} M. Sands, \emph{The Physics of Electron Storage Rings: An Introduction}.  SLAC-R-121, University of California, Santa Cruz, California, USA, 1970 and revised, 1979, available at the URL \url{http://www-public.slac.stanford.edu/scidoc/docMeta.aspx?slacPubNumber=slac-R-121}.
	
	\bibitem{bib:gerke1974} H. Gerke, W. Quarz, \emph{Cavity resonators for a 3\,GeV double-ring storage facility}. Kerntechnik 6 (1974), p.~246-251.
	
	\bibitem{bib:wille2000} K. Wille, \emph{The Physics of Particle Accelerators}, Oxford Univ. Press, Oxford, 2000.
	
	\bibitem{bib:ratzinger2005} U. Ratzinger, \emph{H-type linac structures}. In: CERN Accelerator School: Radio Frequency Engineering, Seeheim, Germany, 2005, available at the URL \url{http://cds.cern.ch/record/865926}.
	
	\bibitem{bib:padamsee2008} H. Padamsee, J. Knobloch, T. Hays, \emph{RF Superconductivity for Accelerators}, Wiley, Weinheim, 2nd edition, 2008.
	
	\bibitem{bib:padamsee2013} H. Padamsee, \emph{Design Topics for Superconducting RF Cavities and Ancillaries}. In: CERN Accelerator School: Superconductivity for Accelerators, Erice, Italy, 2013, available at the URL \url{http://cds.cern.ch/record/1974051}.
	
	\bibitem{bib:xiao2015} B. Xiao et al., \emph{Design, prototyping, and testing of a compact superconducting double quarter wave crab cavity}. Phys. Rev. ST Accel. Beams \textbf{18}, p.~041004, 2015, available at the URL \url{https://journals.aps.org/prab/abstract/10.1103/PhysRevSTAB.18.041004}.
    
    \bibitem{bib:gardner2000} I.~S.~K.~Gardner, \emph{Ferrite dominated cavities}. In CERN Accelerator School: Radio Frequency Engineering, Seeheim, Germany, 2000, available at the URL \url{http://cds.cern.ch/record/400743}.
    
    \bibitem{bib:klingbeil2010} H. Klingbeil, \emph{Ferrite cavities}. In CERN Accelerator School: RF for Accelerators, Ebeltoft, Denmark, 2010, available at the URL \url{http://cds.cern.ch/record/1411778}. 
    
    \bibitem{bib:mori1998} Y. Mori et al., \emph{A New Type of RF Cavity for High Intensity Proton Synchrotron using High Permeability Magnetic Alloy}. EPAC1998, Rome, Italy, 1998, pp.~299--301, available at the URL \url{http://www.cern.ch/accelconf/e98/PAPERS/THOB03B.PDF}.
    
    \bibitem{bib:hanna1993} S.~M. Hanna, P.~M. Stefan, \emph{Application of impedance measurement techniques to accelerating cavity mode characterization}, Nucl. Instr. and Methods A \textbf{335}, 1993, pp.~367--376, available at the URL \url{https://doi.org/10.1016/0168-9002(93)91220-H}.

    \bibitem{bib:jensen2010} E. Jensen, \emph{Cavity basics}. In CERN Accelerator School: RF for Accelerators, Ebeltoft, Denmark, 2010, available at the URL \url{http://cds.cern.ch/record/1416619}.
    
    \bibitem{bib:jensen2013} E. Jensen, \emph{RF Cavity Design}.  CERN Accelerator School: Advanced Accelerator Physics, Trondheim, Norway, 2013, available at the URL \url{http://cds.cern.ch/record/1982429}.
    
    \bibitem{bib:boussard1999} D. Boussard, T. Linnecar, \emph{The LHC Superconducting RF System}. CERN-LHC-Project-Report-316, CERN, Geneva, Switzerland, 1999, available at the URL \url{http://cds.cern.ch/record/410377}.
    
    \bibitem{bib:boussard1993} D. Boussard, V. R\"odel, \emph{The LHC RF System}, 15th International Conference on High-Energy Accelerators, Int. J. Mod. Phys. A, Proc. Suppl. 2B, 1993, pp.~754--756, available at the URL \url{http://cds.cern.ch/record/239611}.
    
    \bibitem{bib:alesini2010} D. Alesini, \emph{Power coupling}. CERN Accelerator School: RF for Accelerators, Ebeltoft, Denmark, 2010, available at the URL \url{http://cds.cern.ch/record/1407380}.
    
    \bibitem{bib:haebel1996} E. Haebel, \emph{Couplers for Cavities}. CERN Accelerator School: Superconductivity in Particle Accelerators, Hamburg, Germany, 1996, available at the URL \url{http://cds.cern.ch/record/308016/}.
    
    \bibitem{bib:garoby1997} R. Garoby et al., \emph{The PS 40 MHz Bunching Cavity}. PAC1997, Vancouver, B.C., Canada, 1997, pp.~2953--2955, available at the URL \url{https://accelconf.web.cern.ch/pac97/papers/pdf/2P036.PDF}.
    
    \bibitem{bib:kindermann1999} H.-P. Kindermann, M. Stirbet, \emph{The variable power coupler for the LHC superconducting cavity}. 9th Workshop on RF superconducting cavity, Santa Fe, New Mexico, USA, 1999, available at the URL \url{http://cds.cern.ch/record/420403}.
    
    \bibitem{bib:boussard1991} D. Boussard, \emph{RF Power Requirements for a High Intensity Proton Collider}. CERN-SL-91-16-RFS, CERN, Geneva, Switzerland, 1991, available at the URL \url{http://cds.cern.ch/record/221153}
    
    \bibitem{bib:baudrenghien2007} P. Baudrenghien, \emph{The Tuning Algorithm of the LHC 400 MHz Superconducting Cavities}. CERN-AB-Note-2007-011, CERN, Geneva, Switzerland, 2007, available at the URL \url{http://cds.cern.ch/record/1014303}.
    
    \bibitem{bib:mastoridis2017} T. Mastoridis, P. Baudrenghien, J. Molendijk, \emph{Cavity voltage phase modulation to reduce the high-luminosity Large Hadron Collider rf power requirements}. Phys. Rev. Accel. Beams \textbf{20}, p.~101003, 2017, available at the URL \url{https://journals.aps.org/prab/abstract/10.1103/PhysRevAccelBeams.20.101003}.
    
    \bibitem{bib:montesinos2018} E. Montesinos, \emph{Accelerators for Medical Applications - Radio Frequency Powering}. In CERN Accelerator School: 	Accelerators for Medical Applications, V\"osendorf Austria, 2018, available at the URL \url{http://cds.cern.ch/record/2315172}.
    
    \bibitem{bib:siemens1994} Siemens AG, Bereich Passive Bauelement und R\"ohren, \emph{Transmitter and Generator Tubes}. Siemens AG, Munich, Germany, 1994, available at the URL \url{https://frank.pocnet.net/other/Siemens/sende-und_generatorroehren_199409_iss3.pdf}.
    
    \bibitem{bib:wood2008} J.~Wood et al., \emph{Characterization and Modeling of LDMOS Power FETs for RF Power Amplifier Applications}, 2008 IEEE Topical Meeting on Silicon Monolithic Integrated Circuits in RF Systems,  Orlando, Florida, USA, 2008, available at the URL \url{https://ieeexplore.ieee.org/document/4446274}.
    
    \bibitem{bib:dye1992} N. Dye and H. Granberg, \emph{Radio Frequency Transitors}, Butterworth-Heinemann, Stoneham, Massachusetts, 1992.
    
    \bibitem{bib:jacob2010} J.~Jacob, \emph{Soleil Experience with High Power Solid State Amplifiers}. RFTEch 2nd Workshop, PSI, Villigen, Switzerland, available at the URL \url{https://lpsc-indico.in2p3.fr/event/530/contributions/3356/}.
    
    \bibitem{bib:jacob2016} J.~Jacob, \emph{Radio Frequency Solid State Amplifiers}. In CERN Accelerator School: Power Converters, Baden, Switzerland, 2014, available at the URL \url{http://cds.cern.ch/record/2038626}.
    
    \bibitem{bib:baudrenghien2005a} P. Baudrenghien, \emph{Low level RF systems for synchrotrons. part II: High Intensity. Compensation of the beam induced effects}. In: CERN Accelerator School: Radio Frequency Engineering, Seeheim, Germany, 2005, available at the URL \url{http://cds.cern.ch/record/702662}.
    
    \bibitem{bib:garoby1991} R. Garoby,
	\emph{Low level RF building blocks}. In: CERN Accelerator School: RF Engineering for Particle Accelerators, Oxford, United Kingdom, 1991, available at the URL \url{http://cds.cern.ch/record/225609}.

	\bibitem{bib:boussard1973} D. Boussard, \emph{Une pr\'{e}sentation \'{e}l\'{e}mentaire du syst\`{e}me Beam Control du PS}. MPS/SR/Note/73-10, CERN, Geneva, Switzerland, 1973, available at the URL \url{http://cds.cern.ch/record/2311314}.
	
	\bibitem{bib:baudrenghien2005b} P. Baudrenghien, \emph{Low-level RF systems for synchrotrons. Part I: The low-intensity case}. In: CERN Accelerator School: Radio Frequency Engineering, Seeheim, Germany, 2005, available at the URL \url{http://cds.cern.ch/record/865920}.
    
    \bibitem{bib:bovet1970} C. Bovet et al., \emph{A selection of formulae and data useful for the design of A.G. synchrontrons}. CERN-MPS-SI-INT-DL-70-4, CERN, Geneva, Switzerland, 1970, available at the URL \url{http://cds.cern.ch/record/104153}.

\end{thebibliography}
\end{document}